\documentclass[]{aastex631}
\renewcommand{\edit}[1]{}

\usepackage{graphicx}
\usepackage{float}
\usepackage[caption=false]{subfig}
\usepackage{enumitem}
\usepackage{comment}

\newcommand{\methane}{CH$_4$}
\newcommand{\cotwo}{CO$_2$}
\newcommand{\cotwoSignificance}{2.4 $\sigma$}
\newcommand{\chfourSignificance}{2.0 $\sigma$}

\newcommand{\ngts}{NGTS}
\newcommand{\tess}{TESS}

\shorttitle{Detection of GJ 1214 b Molecules}
\shortauthors{Schlawin, Ohno, Bell et al.}

\begin{document}

\title{Possible Carbon Dioxide Above the Thick Aerosols of GJ 1214 b}

\correspondingauthor{Everett Schlawin}
\email{eas342 AT EMAIL Dot Arizona .edu}

\author[0000-0001-8291-6490]{Everett Schlawin}
\affiliation{Steward Observatory, 933 North Cherry Avenue, Tucson, AZ 85721, USA}

\author[0000-0003-3290-6758]{Kazumasa Ohno}
\affil{Division of Science,
National Astronomical Observatory of Japan,
2-12-1 Osawa,
Mitaka-shi 1818588 Tokyo, Japan}

\author[0000-0003-4177-2149]{Taylor J. Bell}
\affiliation{Bay Area Environmental Research Institute, NASA's Ames Research Center, Moffett Field, CA 94035, USA}
\affiliation{Space Science and Astrobiology Division, NASA's Ames Research Center, Moffett Field, CA 94035, USA}

\author[0000-0002-8517-8857]{Matthew M. Murphy}
\affiliation{Steward Observatory, 933 North Cherry Avenue, Tucson, AZ 85721, USA}

\author[0000-0003-0156-4564]{Luis Welbanks}
\affiliation{School of Earth and Space Exploration, Arizona State University, Tempe, AZ, USA}

\author[0000-0002-9539-4203]{Thomas G. Beatty}
\affiliation{Department of Astronomy, University of Wisconsin--Madison, Madison, WI 53703, USA}

\author[0000-0002-8963-8056]{Thomas P. Greene}
\affiliation{Space Science and Astrobiology Division, NASA's Ames Research Center, Moffett Field, CA 94035, USA}

\author[0000-0002-9843-4354]{Jonathan J. Fortney}
\affiliation{Department of Astronomy and Astrophysics, University of California, Santa Cruz, CA, USA}

\author[0000-0001-9521-6258]{Vivien Parmentier}
\affiliation{Laboratoire Lagrange, Observatoire de la Côte d’Azur, CNRS, Université Côte d’Azur, Nice, France}

\author[0000-0001-8745-2613]{Isaac R. Edelman}
\affiliation{Bay Area Environmental Research Institute, NASA's Ames Research Center, Moffett Field, CA 94035, USA}

\author[0000-0002-4259-0155]{Samuel Gill}
\affiliation{Centre for Exoplanets and Habitability, University of Warwick, Gibbet Hill Road, Coventry CV4 7AL, UK}
\affiliation{Department of Physics, University of Warwick, Gibbet Hill Road, Coventry CV4 7AL, UK}

\author[0000-0001-7416-7522]{David R. Anderson}
\affiliation{Instituto de Astronom\'ia, Universidad Cat\'olica del Norte, Angamos 0610, 1270709, Antofagasta, Chile}

\author[0000-0003-1452-2240]{Peter J. Wheatley}
\affiliation{Centre for Exoplanets and Habitability, University of Warwick, Gibbet Hill Road, Coventry CV4 7AL, UK}
\affiliation{Department of Physics, University of Warwick, Gibbet Hill Road, Coventry CV4 7AL, UK}

\author[0000-0003-4155-8513]{Gregory W. Henry}
\affiliation{Tennessee State University, Nashville, TN  37209, USA (retired)}

\author[0000-0001-6086-4175]{Nishil Mehta}
\affiliation{Université Côte d'Azur, Observatoire de la Côte d'Azur, CNRS, Laboratoire Lagrange, France}

\author[0000-0003-0514-1147]{Laura Kreidberg}
\affiliation{Max-Planck-Institut für Astronomie, Königstuhl 17, D-69117 Heidelberg, Germany}

\author[0000-0002-7893-6170]{Marcia J. Rieke}
\affiliation{Steward Observatory, 933 North Cherry Avenue, Tucson, AZ 85721, USA}



\begin{abstract}

Sub-Neptune planets with radii \edit1{smaller than} Neptune (3.9 R$_\oplus$) are the most common type of planet known to exist in The Milky Way, even though they are absent in the Solar System.
These planets can potentially have a large diversity of compositions as a result of different mixtures of rocky material, icy material and gas accreted from a protoplanetary disk.
However, the bulk density of a sub-Neptune, informed by its mass and radius alone, cannot uniquely constrain its composition; atmospheric spectroscopy is necessary.
GJ 1214 b, which hosts an atmosphere that is potentially the most favorable for spectroscopic detection of any sub-Neptune, is instead enshrouded in aerosols (thus showing no spectroscopic features), hiding its composition from view at previously observed wavelengths in its terminator.
Here, we present a JWST NIRSpec transmission spectrum from 2.8 to 5.1~\micron\ that shows signatures of CO$_2$ and CH$_4$, expected at high metallicity.
A model containing both these molecules is preferred by 3.3 and 3.6~$\sigma$ as compared to a featureless spectrum for two different data analysis pipelines, respectively.
Given the low signal-to-noise of the features compared to the continuum, however, more observations are needed to confirm the CO$_2$ and CH$_4$ signatures and better constrain other diagnostic features in the near-infrared.
Further modeling of the planet's atmosphere, interior structure and origins will provide valuable insights about how sub-Neptunes like GJ 1214~b form and evolve.

\end{abstract}

\keywords{Exoplanet atmospheric composition --- Exoplanet atmospheres -- Exoplanet astronomy --- Exoplanet atmospheric structure}



\section{Introduction} \label{sec:intro}

Observations with Kepler and TESS have shown that planets are very common in the Milky Way, with an increased frequency of planets with  radii smaller than Neptune (3.9 R$_\oplus$)   \citep{howard2012occurrenceKepler,fulton2017radiusGap,dattilo2023unifiedTreatmentKeplerOccurence}.
With no analog in the solar system between 1.0~R$_\oplus$ and 3.9~R$_\oplus$, these planets' nature is of high interest and is still largely unknown \citep[e.g.,][]{miller-ricci2010natureOfGJ1214b} and they have been dubbed as Super-Earths, mini Neptunes and sub-Neptunes \citep[e.g][]{charbonneau2009gj1214bDiscovery,benneke2012retrieval,bean2021subNeptunesReview}.
Despite their prevalence, the major compositions of super-Earths and sub-Neptunes remain elusive.
These planets' densities, informed by their radial-velocity-derived masses and transit-derived radii alone, still allow a wide variety of possible compositions \citep{adams2008OceanVsThickAtm,rogers2010threeOriginsGJ1214b,valencia2013bulkCompositionGJ1214b}. 

Transmission spectroscopy, where the light from the host star is filtered through the planet's atmosphere during its passage in front of the star, offers a way to probe the chemical composition of the atmosphere \citep{kempton2024transitingAtmospheresJWST}.
The unprecedented precision and wavelength coverage of JWST has recently revealed that some sub-Neptune atmospheres have ${\sim}100-200\times$ solar metallicity \citep{madhusudhan2023k2d18,beatty24gj3470b,benneke2024toi270dMiscibleMetalRich,holmberg2024toi270d}, which is comparable to or even higher than ${\sim}80\times$ solar metallicity estimated for Uranus and Neptune in our Solar System \citep{Atreya2020icegiant}.
The diversity of sub-Neptunes atmospheric properties and their relation to the planet's interior are of great interest because they give clues about how and where these planets are formed \citep{benneke2013volatileVsCloudyMiniNeptunes,bean2021subNeptunesReview}.

The statistics of transiting planets in the sub-Neptune size regime reveals a gap in the occurrence rate of planets between 1.5 to 2.0 $R_\oplus$ \citep{fulton2017radiusGap}.
Models can explain this gap as a natural occurrence of the atmospheric loss of massive planets that either retain or lose their Hydrogen/Helium envelopes, whether it is by photoevaporation \citep{owen2017evaporationValley} or core-powered mass loss \citep{gupta2019MNRAS.487...24G}.
The location of the gap with semimajor axis disfavors a preponderance of ``water-worlds'' for FGK type host stars and instead prefers compositions that have an Earth-like, rocky core with a H$_2$-He atmosphere \citep{bean2021subNeptunesReview,gupta2019MNRAS.487...24G,owen2017evaporationValley,wu2019masScalingSuperE,ginzburg2018corePoweredMLRadDistribution}.
However, the planet formation process may create a variety of compositions of sub-Neptunes.
Planets that orbit M-dwarf stars appear to have a clustering in density indicating possible water-rich sub-Neptunes near half the density of Earth \citep{luque2022densityNotRadius,venturini2024fadingRadiusValleyMdwarfs}.

GJ 1214 b \citep{charbonneau2009gj1214bDiscovery} is a benchmark sub-Neptune with a zero-albedo equilibrium temperature of $\sim$600 K, a radius of 2.7~R$_\oplus$, and a mass of 8.2~M$_\oplus$ orbiting a 0.18 M$_\odot$, 0.22 R$_\odot$ M4 type star - see \citet{cloutier2021gj1214bPreciseMass} and \citet{Mahajan+24_gj1214update}.
Given its low bulk density, the planet must have an atmosphere, while the density can be consistent with various interior compositions, including a rocky core with a hydrogen-rich envelope and a nearly pure H$_2$O planet \citep{valencia2013bulkCompositionGJ1214b,rogers2010threeOriginsGJ1214b,Nettelmann2011gj1214}.
While early studies of GJ 1214~b described it as a super-Earth \citep{charbonneau2009gj1214bDiscovery}, its radius is above the 1.5 to 2.0 ~$R_\oplus$ radius gap \citep{fulton2017radiusGap}, so it is more commonly now described as a sub-Neptune \citep{Nixon+24internalStructureGJ1214bJWST,bean2021subNeptunesReview}.
GJ 1214 b has been observed extensively from ground-based and space-based observatories due to its low density, small radius host star and nearby location (14.6 pc), which give the planet the highest Transmission Spectroscopy Metric \citep{kempton2018TESSmetric} of any sub-Neptune and enables investigation into its compositional nature from transmission spectroscopy. 
However, GJ 1214 b's transmission spectrum has been remarkably flat, (i.e.\ featureless) as compared to aerosol-free atmospheres \citep{Bean10,desert2011gj1214bMetalRich,berta2012flat_gj1214,colon2013gj1214bNarrowband,wilson2014gj124bNarrowband,deMooij2012gj1214bOpticalNIR,nascimbeni2015gj1214bLBT,gao2023gj1214} and even aerosol-free metal-rich ones dominated by CH$_4$, H$_2$O or CO$_2$ \citep{kreidberg14}.
The flat, mostly featureless spectrum is explained by high-altitude aerosols or photochemical hazes formed in the upper atmosphere \citep{morley2015superEclouds,KawashimaIkoma2018_haze,Adams+19aggregateHazes,Lavvas+19photochemicalHazes,gao2023gj1214} or KCl clouds lifted up from the lower atmosphere \citep{Charnay+15GJ1214bspectrum,Ohno&Okuzumi18microphysicalCloudModelsGJ214gj436b,gao2018microphysicsGJ1214clouds,Ohno+20fluffyaggregate}.
Although GJ~1214~b has served as a benchmark system for studying aerosol formation in exoplanets for a decade, its aerosol's nature and relations to the planet's background atmospheric composition remain elusive.

A recent breakthrough has enabled some constraints on the composition of its atmosphere and aerosols. \citet{kempton2023reflectiveMetalRichGJ1214} used a full orbit phase curve with JWST MIRI LRS to constrain the Bond Albedo at 0.51$\pm$0.06.
The relatively small day-night contrast compared with general circulation models, constrains the atmospheric metallicity to be more than 100~$\times$ solar.
Furthermore, the dayside and nightside emission spectra of the planet show signatures of molecular absorption at a 3$\sigma$ confidence level, expected from H$_2$O in the MIRI LRS bandpass.
Thus, JWST is finally beginning to enable insights into the nature of GJ 1214 b.

Here, we report the results of two new transmission spectroscopy observations with JWST NIRSpec that shed light on the key atmospheric features in the 2.8 to 5.1~\micron\ wavelength region.
Section \ref{sec:obs} describes the observations and data analysis, and Section \ref{sec:results} describes the model fits to the spectrum.
Finally, we conclude in Section \ref{sec:conclusions}.
The appendix includes details about the observing specifications (Section \ref{sec:obsSpecifications}), transmission spectrum data analysis (Section \ref{sec:dataAnalysis}) and atmospheric models (Section \ref{sec:method_model}).
The appendix also includes a stellar variability analysis (Section \ref{sec:stellarVariability}), Bayesian comparisons between a flat line and \edit1{a} metal-dominated atmospheric model (Section \ref{sec:bayesFlatLine}) and models of stellar spot heterogeneities (Section \ref{sec:spotContamination}).

\section{Observations and Analysis}\label{sec:obs}

We observed the GJ 1214 system with JWST NIRSpec's G395H Bright Object Time Series mode during two consecutive transits beginning exposures on UT 2023-07-18 and 2023-07-19 as part of the MANATEE program (GTO-1185) \citep[e.g.,][]{schlawin2018JWSTforecasts}.
The stellar integrated counts averaged over all integrations were consistent to within 0.2\% between the two transit observations, indicating no significant stellar activity differences between the observations.
Furthermore, long-term ground-based monitoring of the \edit1{GJ 1214} system showed no significant drops in brightness during the epochs of observation and that the overall flux was near the upper range of a long-term monitoring campaign, indicating less overall spot coverage -- see Section \ref{sec:stellarVariability}.

We calculated transmission spectra of GJ 1214 b with 2 different data analysis codes: \texttt{tshirt}\footnote{\url{https://github.com/eas342/tshirt/}} and Eureka!\ \citep{bell2022eureka}, 
similar to those used on previous MANATEE targets \citep{bell2023_methane,beatty24gj3470b,welbanks2024wasp107b}.
We first used broadband light curves from 2 JWST NIRSpec transits, 1 JWST MIRI transit and 2 JWST MIRI eclipses from a previously published phase curve \citep{kempton2023reflectiveMetalRichGJ1214} to derive orbital parameters, listed in Table \ref{exttab:orbparams} for spectroscopic analysis.
The light curve extractions beginning with un-calibrated data and broadband orbital parameter fits are described in Section \ref{sec:tshirtExtraction} through \ref{sec:orbParam}.
We fit the NIRSpec spectroscopic light curves at a constant resolving power of $R$=100, using a scaled Phoenix limb darkening model \citep{grant2024exoticLDjoss,husser2013phoenix} -- see Sections \ref{sec:tshirtLCfitting} and \ref{sec:eurekaLCfitting}.
We include a linear baseline trend with time (for each exposure) and fit for the transit depth as a function of wavelength for the two consecutive transits and calculate a weighted average of the two transmission spectra.
We also re-fitted the MIRI LRS full-orbit spectroscopic phase curve of \citet{kempton2023reflectiveMetalRichGJ1214} with \texttt{Eureka!}, using the same fixed orbital parameters from Table \ref{exttab:orbparams} and applied the lessons learned from the analysis of the Early Release Science phase curve of WASP-43b presented by \citet{bell2024nightsideCloudsDisEqChemWASP43b} -- see Sections \ref{sec:eurekaExtraction} and \ref{sec:eurekaLCfitting}.
Compared to the previously published transit spectrum \citep{kempton2023reflectiveMetalRichGJ1214}, our re-fitted MIRI LRS transmission spectrum was slightly lower by a median of 29 ppm (less than 1$\sigma$ change) and had marginally larger error bars (40\% larger on the median).
Finally, we re-fit the broadband light curves from HST scan mode observations \citep{kreidberg14} with these new orbital parameters to derive an overall offset to the spectrum from \citet{kreidberg14} downward by 119 ppm -- see Section \ref{sec:hstOffset}.

\begin{figure*}
    \includegraphics[width=0.98\linewidth]{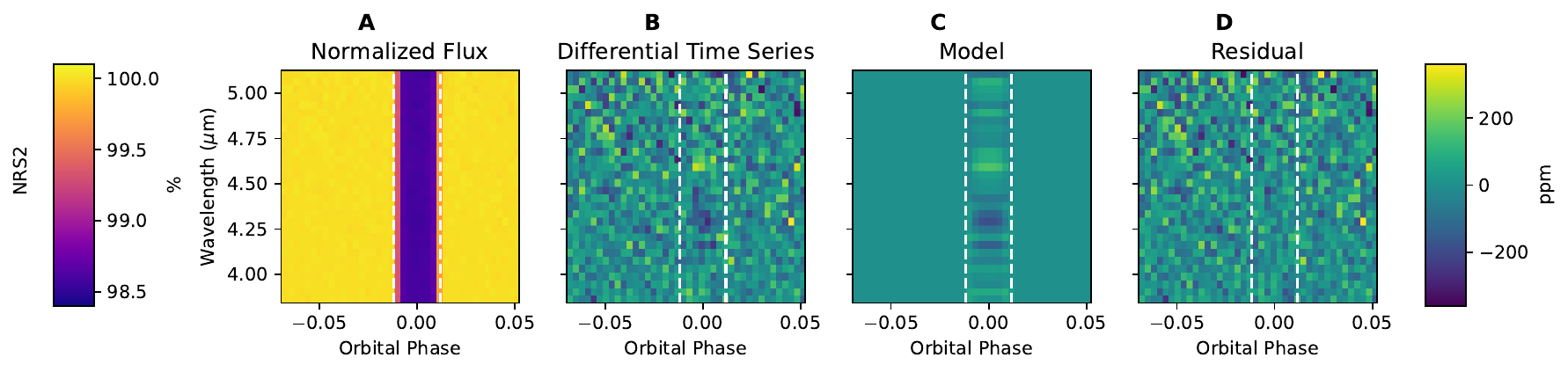}
    \includegraphics[width=0.98\linewidth]{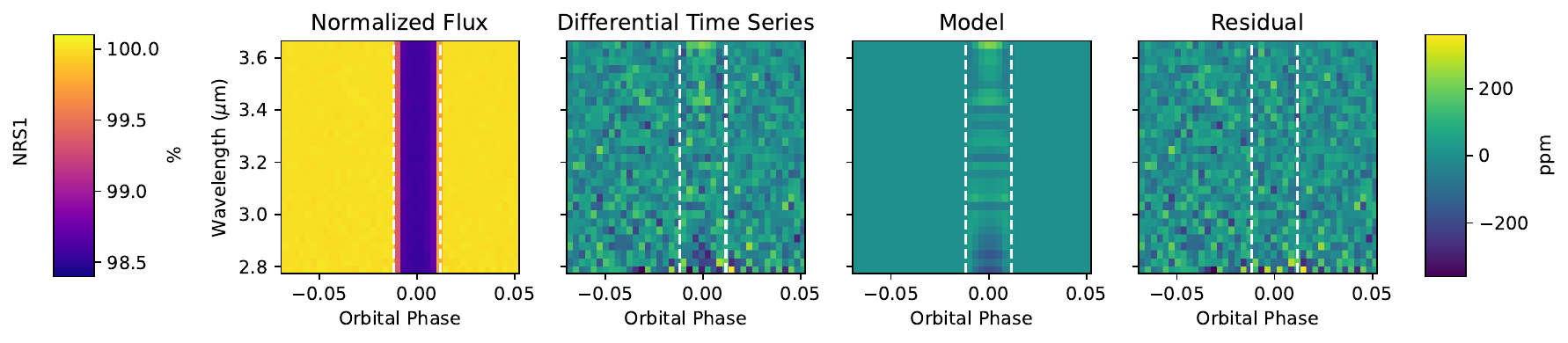}
    \caption{2D light curves binned to a resolution of R=100 and 40 time bins for NRS2 (top row) and NRS1 (bottom row), with both transits averaged. The normalized flux in percentage (after linearly de-trending at each wavelength) is shown in panel A. The differential time series, after dividing by the broadband for a given detector is shown in ppm in panel B. The differential light curve model is shown in panel C and the residuals of the light curve subtracted by the model are shown in panel D.
    The start of ingress and end of egress times are marked as vertical dashed lines.
    Increased and decreased transit depths relative to the broadband light curve are visible as dark and bright bands between ingress and egress.
    The most noticeable features are increased transit depths near 4.3~\micron\ and 2.8~\micron\ and decreased transit depths near 3.6~\micron\ and 4.6~\micron.}
    \label{fig:2Dlc}
\end{figure*}

Our JWST NIRSpec observations, differential time series and residuals are shown in 2D form in Figure \ref{fig:2Dlc}, where both transits are co-phased and binned to 40 evenly-spaced time bins across the 4.6 hours of observation.
The top row is the NRS2 detector, and the bottom row is the NRS1 detector.
Panel A (``Normalized Flux'') shows the overall transit light curve normalized by the out-of-transit flux and linearly de-trended per wavelength channel.
The transit is visible as a vertical dark band.
The times of first contact and fourth contact \citep[e.g.,][]{winnchap} are marked as vertical dashed lines.
We next divide each detector time series by the average/white light curve (i.e. the average value per time bin) to create a differential time series, which better shows spectral changes during transit as seen in Figure \ref{fig:2Dlc}'s panel B (``Differential Time Series'').
In panel B, the differential time series shows decreased flux (extra absorption by the planet) near 4.3~\micron\ on NRS2 and near 2.8~\micron\ on NRS1.
We fit the light curves of GJ 1214 b at full time resolution and individually per transit observation with a linear baseline (Described in Sections \ref{sec:tshirtLCfitting} and Section \ref{sec:eurekaLCfitting}).
We fit the light curve models with \texttt{pymc3} \citep{salvatier2016pymc3} using a \texttt{starry} \citep{luger2019starry} light curve model at full resolution.
We also bin the model (with the linear baseline removed) and divide by the average light curve per detector as shown in Figure \ref{fig:2Dlc}'s panel C (``Model'').
In other words, the images in panel B (``Differential Time series'') and panel C (``Model'') are calculated the same way but one is for the observed data and the other is for the light curve models evaluated at the same orbital phases.
Finally, we calculate the light curve residuals by subtracting the light curves by the wavelength-dependent \texttt{starry} model and show the residual in Figure \ref{fig:2Dlc}'s Panel D (``Residual'').

As visible in the 2D light curves and residuals, the largest deviations are during the transit between first and fourth contact and show dark bands near 2.8~\micron\ on NRS1 and 4.3~\micron\ in NRS2 (thus showing decreased flux or deeper transits by the planet).
CO$_2$, a prominent gas detected in hot Jupiter and hot Neptune atmospheres \citep[e.g.,][]{ers2023wasp39b_CO2}, has characteristic opacity features near 2.8~\micron\ and 4.3~\micron.
Furthermore, previous metal-rich models for GJ 1214 b predict CO$_2$ in the planet's atmosphere \citep{gao2023gj1214}.
Thus, there are potentially molecular features visible for the first time in the transmission spectrum of the planet, despite the lack of features at 1.1 to 1.7~\micron\ and 5.0 to 11~\micron\ \citep{berta2012flat_gj1214,kreidberg14,kempton2023reflectiveMetalRichGJ1214,gao2023gj1214}.
We next compare the observed spectrum to some illustrative models in Section \ref{sec:results}.

\section{Transmission Spectrum Results}\label{sec:results}

\begin{figure}
    \centering
    \includegraphics[width=0.75\linewidth]{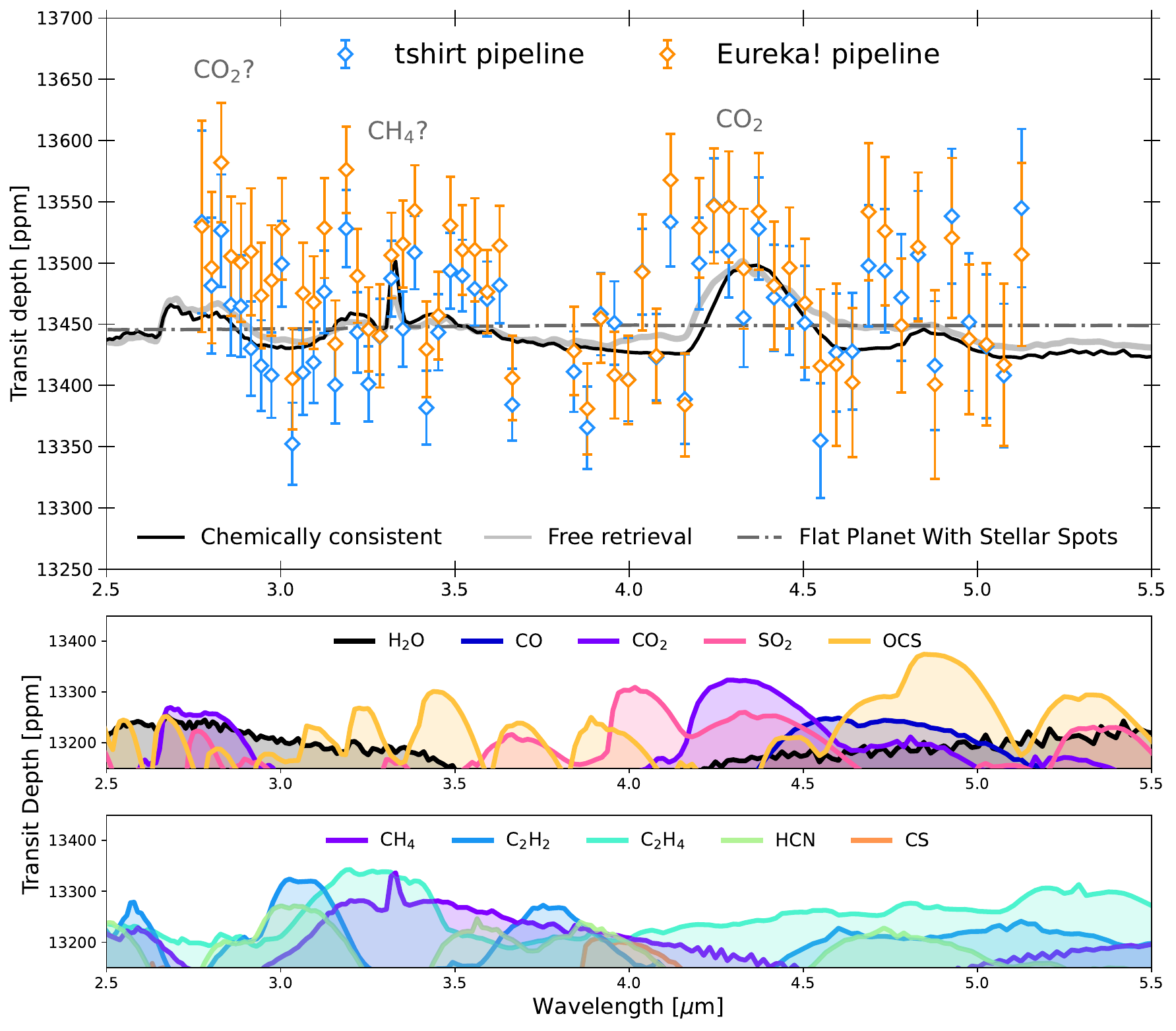}
    \caption{\textbf{Transmission Spectrum of GJ 1214 b with Two Data Reductions and Illustrative Models:}
    The top panel shows the results of data reduction by the \texttt{tshirt} and \texttt{Eureka!} pipelines, demonstrating that two independent pipelines yield consistent results within $1\sigma$ errorbars.
    The solid black, solid gray and dash-dotted gray lines show the median spectrum from the chemically-consistent model, chemically-agnostic free model, and a flat line (with stellar inhomogeneities), respectively fit to the HST, JWST-NIRSpec (\texttt{tshirt}) and JWST-MIRI panchromatic spectrum.
    The lower panels show the illustrative transmission spectra to exhibit the transit depth shape of relevant molecules -- we create clear isothermal ($T=500~{\rm K}$)  N$_2$ atmospheres with only one gas at a time with a VMR=$10^{-4}$ and repeat for all molecules.}
    \label{fig:differentAnalyses}
\end{figure}

Figure \ref{fig:differentAnalyses} shows that the observed spectrum deviates from a flat line (ie. a featureless spectrum) at the 50-80 ppm level.
This was predicted by previous haze models at high (1000$\times$) atmospheric metallicity \citep{gao2023gj1214}.
The observed spectrum shows a noticeable bump around $\sim$4.3~\micron\ and a smaller one at $\sim$2.8~\micron, which are consistent with the characteristic absorption bands of CO$_2$ predicted by previous modeling studies \citep{gao2018microphysicsGJ1214clouds,Lavvas+19photochemicalHazes,Ohno+20fluffyaggregate,gao2023gj1214}.
We confirm that the possible CO$_2$ feature appears in the spectrum reduced by both \texttt{tshirt} and \texttt{Eureka!} as well as the spectrum observed at each individual observational epoch (Section \ref{sec:bayesFlatLine}).
Previous modeling studies also predicted a CH$_4$ feature at $\sim$3.3~\micron\ \citep{miller-ricci2010natureOfGJ1214b,Ohno+20fluffyaggregate}.
The observed spectrum indeed shows modulation around ${\sim}3.3~{\rm {\mu}m}$, consistent with the presence of CH$_4$.
We note that there is no characteristic H$_2$O absorption near 2.8-3.5~\micron, which would be expected for a pure steam atmosphere as hypothesized to exist around some sub-Neptunes like GJ 1214~b \citep{miller-ricci2010natureOfGJ1214b,valencia2013bulkCompositionGJ1214b,luque2022densityNotRadius,deMooij2012gj1214bOpticalNIR,kempton2023waterWorlds}.

We further interpret the spectrum of GJ1214b by inferring its atmospheric properties using a Bayesian inference algorithm.
We use Aurora \citep{Welbanks&Madhusudhan21Aurora}, a retrieval framework that can explore different atmospheric compositions beyond the usual H-rich composition for gas giants including H$_2$O rich atmospheres (e.g., steam atmospheres), N$_2$ rich atmospheres, and CO$_2$ rich atmospheres. 
Our model considers the possibility of spectroscopic signatures and contributions to the atmospheric mean molecular weight from 
a mixture of H$_2$ and He following a solar He/H$_2$ ratio of 0.17 \citep{Asplund2009}, H$_2$O, CH$_4$, N$_2$, CO$_2$, CO, HCN, NH$_3$, OCS, CS$_2$, H$_2$S, SO$_2$, C$_2$H$_2$ and PH$_3$ (i.e. 14 abundances). 
Further details about the model and parameter posterior distributions are in Section \ref{sec:freeRetrieval}.

We perform a Bayesian model comparison to assess the model preference for these CO$_2$ and CH$_4$ signatures \citep[i.e., nested retrievals, see e.g., ][]{Welbanks2023}.
The presence of \cotwo\ in our 29 parameter model is preferred at \cotwoSignificance\ while the presence of \methane\ is preferred at \chfourSignificance.
Given the signal-to-noise of the observations, we prefer to use this more agnostic and conservative modelling approach with a large number of parameters considering any possible background gas rather than constructing a simpler model that may yield higher model preferences (e.g., ``detection significances'').
The large number of parameters also allows us to check whether other molecules or cloud parameters may mimic the observed features and noise in the spectrum and thus decrease the detection significance.
Figure \ref{fig:differentAnalyses} shows the median retrieved atmospheric model showing the spectroscopic features of \cotwo\ and \methane\ in the NIRSpec G395H bandpass. The inferred presence of clouds and hazes and a high-mean molecular weight atmosphere (i.e., low atmospheric scale height) results in a featureless spectrum previously observed in the HST and MIRI bands -- see Section \ref{sec:panchromatic} and Figure \ref{fig:panchromaticSpec}.

For a cross-check, we also performed a chemically-consistent model fit using an analytical Temperature-Pressure profile of \citet{guillot2010radEquilibrium} with parameterized aerosol opacity and thermochemical equilibrium compositions supplemented by a disequilibrium quenching prescription (see Section \ref{sec:chemConsistentRetrieval} for detail). 
This second model also fits the 4.3~\micron\ and 2.8~\micron\ features with \cotwo, shows \methane\ absorption near 3.3~\micron\ and a weak OCS feature.
The OCS feature is not strong, so there is no preference for the gas over removing it in the chemically-free model described above.

We consider whether the observed transit features are purely due to unocculted stellar spots or plages that have different spectra than the surrounding photosphere \citep[e.g.,][]{pont2008hazeHD189733b,sing2011hstHD189733b,rackham2018transitSourceEffect} and that the planet spectrum is featureless.
We model the stellar photosphere with \texttt{SPHINX} \citep{Iyer+23sphinx} with a constant 0.032 spot coverage fraction -- see Section \ref{sec:spotContamination}.
The best fit of the spotted star model is nearly the same as a flat line because the broad 2.8-3.4~\micron\ stellar feature predicted by \texttt{SPHINX} \citep[which can mimic H$_2$O features in the planet's atmosphere, e.g.,][]{moran2023gj486bwater} is not present in our NIRSpec spectrum -- see Figure \ref{fig:differentAnalyses} and Section \ref{sec:spotContamination} for more details.
The CO$_2$ features at 2.8~\micron\ and 4.3~\micron\ do not correspond to known stellar inhomogeneity effects at the spot temperatures expected for GJ 1214 A.

\begin{figure}
    \centering
     \includegraphics[width=0.49\linewidth]{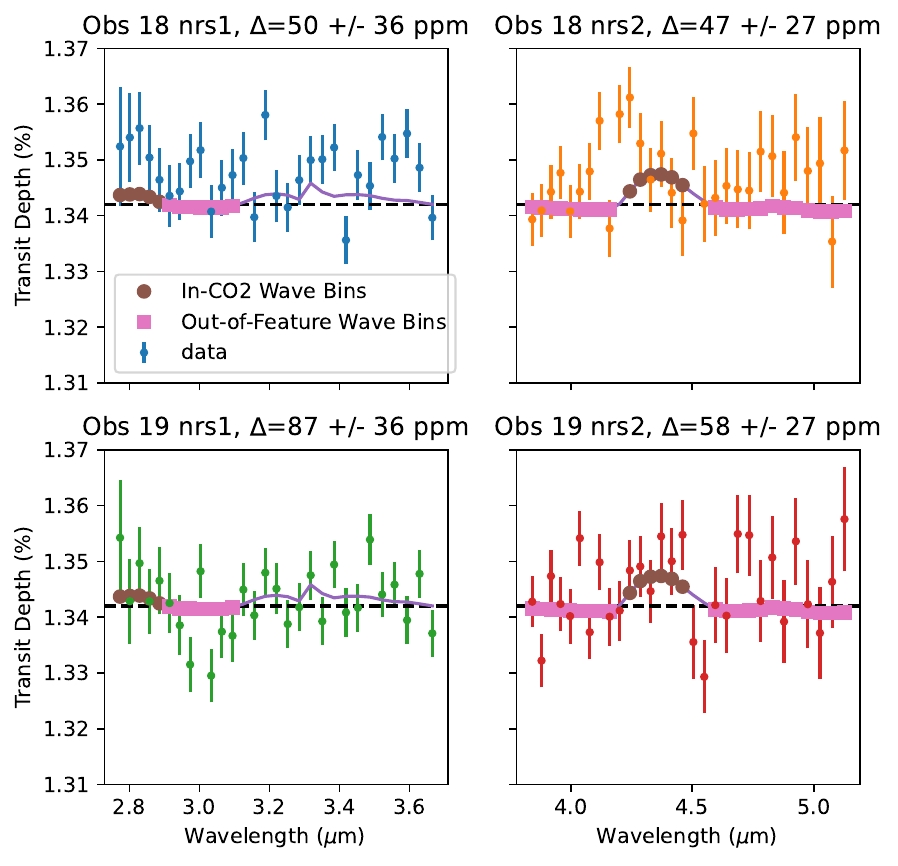}
    \includegraphics[width=0.49\linewidth]{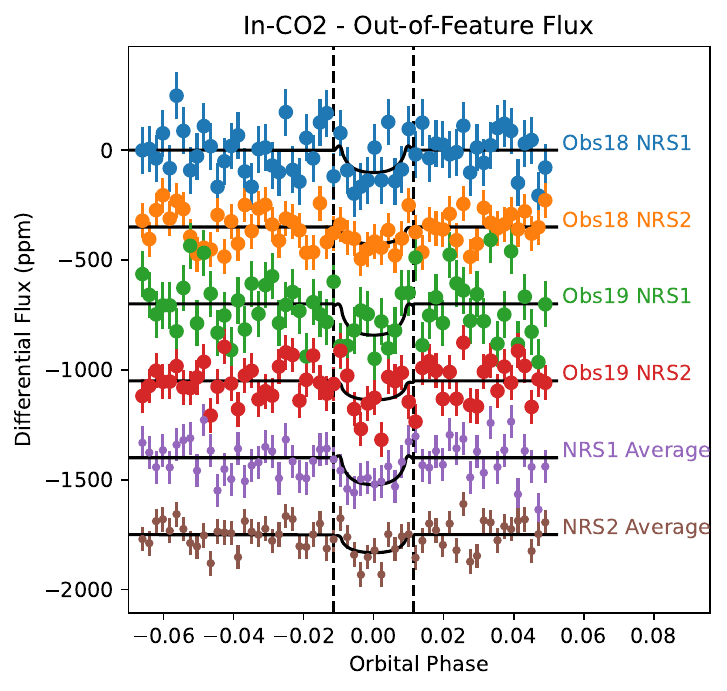}
    \caption{{\it Left:} Wavelength selections for CO$_2$ and Out-of-Feature co-adding.
    We use the chemically-consistent model (purple curve) and a threshold (dashed black line) to select wavelengths that are dominated by CO$_2$ opacity (brown circles) and wavelengths that are outside of features (pink circles).
    Individual spectra are shown for the NRS1 and NRS2 detectors and per observation (Obs 18 NRS1 (blue), Obs 18 NRS2 (orange), Obs 19 NRS1 (green), and Obs 19 NRS2 (red)).
    {\it Right:} Differential light curve inside and outside of significant CO$_2$ features. Each detector and observation's fluxes are summed for wavelengths inside a feature and divided by the fluxes summed outside this feature.
    The same blue, orange, green and red colors are used on the left and right to code each detector and observation.
    Finally, an average in-feature versus out-of-feature light curve across both observations is shown in purple and brown on the bottom for NRS1 and NRS2, respectively.
    The same wavelengths' best fit models are shown as black curves, which use a limb-darkened \texttt{starry} model.
    Linear trends have been removed from all time series and \edit1{vertical offsets were applied} for clarity.}
    \label{fig:CO2lc}
\end{figure}

As another method to visualize the significance of the CO$_2$ detection, we add together the flux at wavelengths both inside the two CO$_2$ features as well as outside of major features at the two transit epochs separately.
We calculate the differential light curves by dividing the time series inside the CO$_2$ feature by the out-of-feature flux and show the resulting light curve in Figure \ref{fig:CO2lc}.
The light \edit1{curves} show absorption at the in-\cotwo\ wavelengths \edit1{in} both detectors and both observations.
We note that the NRS1 \cotwo\ depth appears larger due to limb darkening, but the 4.3~\micron\ \cotwo\ feature is expected to be larger.
The differential light curves show a combined \edit1{weighted average of} 58 $\pm$ 15 ppm  \edit1{transit depth difference (weighted averages of 69 $\pm$ 25 ppm for NRS1 and 53 $\pm$ 19 ppm for NRS2, individually)}.
We explore further tests of the statistical likelihood that GJ 1214 b has a featureless spectrum in JWST NIRSpec in Section \ref{sec:bayesFlatLine}; these tests all prefer a model that contains both CO$_2$ and CH$_4$ \edit1{in the} atmosphere over a featureless flat spectrum.
The two-transit-combined spectrum prefers the model that contains both CO$_2$ and CH$_4$ by 3.3~$\sigma$ and 3.8~$\sigma$ for the \texttt{tshirt} and \texttt{Eureka!} spectra, respectively.

\section{Conclusions and Discussion}\label{sec:conclusions}

We observed GJ~1214~b with JWST NIRSpec on two separate consecutive transits on UT 2023-07-18 and UT2023-07-19.
We used two data analysis pipelines on the uncalibrated data products to extract light curves and fit the light curves to derive a transmissions spectrum.
While previous high precision observations in the near infrared (HST) and mid-infrared (JWST-MIRI) have shown featureless transmission spectra due to thick aerosols \citep{kreidberg14,kempton2023reflectiveMetalRichGJ1214}, the spectrum exhibits \cotwo-like features at 2.8~\micron\ and 4.3~\micron\ in both observations and follows predictions of these moderate features in a highly metal-rich atmosphere \citep{Ohno+20fluffyaggregate,gao2018microphysicsGJ1214clouds}.
We calculated an atmospheric model with both \cotwo\ and \methane\ and find that this model is preferred over a flat line (ie. featureless) spectrum at the 3.3~$\sigma$ and 3.8~$\sigma$ level for the \texttt{tshirt} and \texttt{Eureka!} data analysis pipelines, respectively.
We also performed Bayesian comparisons to assess the model preference for individual gases and find that \cotwo\ is prefererred at \cotwoSignificance\ and that \methane\ is preferred at \chfourSignificance\ for the \texttt{tshirt} data analysis pipeline.
Our derived transmission spectrum does not correlate with any expected stellar spot inhomogeneities with a \texttt{SPHINX} model \citep{Iyer+23sphinx}.
We thus demonstrate the ability of JWST observations to detect molecules even from the most aerosol-rich sub-Neptunes known today.

The relative strength of CO$_2$ and CH$_4$ is a metallicity indicator \citep{moses2013compositionalDiversity,yang2024chemicalMapping}.
The fact that CO$_2$ is more strongly detected than CH$_4$ 
indicates that the planet metallicity is likely very high ($>10^3$ times solar enrichment) \citep{moses2013compositionalDiversity}.
This is similar to the conclusion from haze modeling of the transmission spectrum with HST WFC3 and MIRI LRS \citep{gao2023gj1214}.
Our findings support the high metallicity atmosphere suggested by previous modeling studies \citep{gao2018microphysicsGJ1214clouds,Lavvas+19photochemicalHazes,Ohno+20fluffyaggregate,gao2023gj1214} and by the JWST MIRI phase curve \citep{kempton2023reflectiveMetalRichGJ1214}.
It is also possible for highly enriched atmospheres to more easily form thick clouds, which may be the type of environment responsible for the very high geometric albedo of 0.80 observed in the hot Neptune LTT 9779 b \citep{hoyer2023LTT9779b}.
Meanwhile, such a metal-rich atmosphere may pose a challenge to the mass-radius relation of GJ~1214~b, which requires an extremely high ice fraction within the planet if the atmospheric metallicity is high \citep{Nixon+24internalStructureGJ1214bJWST}.
Further modeling and atmospheric retrievals are warranted to better constrain the chemical composition of GJ~1214~b's observable atmosphere based on its panchromatic transmission spectrum, which are presented in our companion paper (Ohno et al. in prep).

Although our atmospheric observations can provide valuable insights on the interior structures of sub-Neptunes, two transits with JWST NIRspec are too noisy to constrain the chemical inventory of GJ~1214~b in detail.
CO$_2$ and the potential presence of OCS and CH$_4$ provide clues to the atmospheric carbon, oxygen, and sulfur inventory, which may inform us how the planet formed and evolved; however, the detection significance of CO$_2$ and CH$_4$ are only \cotwoSignificance\ and \chfourSignificance\ individually.
OCS is not favored by the free retrieval for current observations, but may exist if the metallicity is indeed high according to previous studies of atmospheric chemistry \citep{moses2013compositionalDiversity,yang2024chemicalMapping}.
\edit1{We also note that the 2.8~\micron\ feature in our observations appears as large or larger than the 4.37~\micron\ \cotwo\ feature, whereas \cotwo\ should have a 4.37~\micron\ transit depth that is 31 ppm larger than the 2.8~\micron\ in our chemically-consistent model.
Complementary observations will help determine if the 2.8~\micron\ feature is affected by systematics at the edge of the NRS1 G395H bandpass, such as with NIRSpec G235H or NIRCam F322W2.}
Further observations are highly warranted to robustly constrain atmospheric elemental abundances in depth and thus better reveal the origin and evolution of GJ~1214~b. 

If the inferred high metallicity of the planet is borne out by future observations, GJ 1214 b would contrast with more moderately enriched sub-Neptunes like K2-18 b \citep{madhusudhan2023k2d18} and TOI-270 d \citep{benneke2024toi270dMiscibleMetalRich,holmberg2024toi270d} and the less metal-enriched GJ 3470 b \citep{beatty24gj3470b}, which could be beyond the runaway accretion mass.
Further searches should also be conducted for signatures of ``water-worlds'' \citep{kempton2023waterWorlds,damiano2024LHS1140bPotentialWaterWorld}, which may be concentrated in some M-dwarf systems \citep{luque2022densityNotRadius,chakrabarty2024whereAreWaterWorlds}.
Understanding GJ~1214~b would also provide insight into the nature of many other sub-Neptunes and super Earths that are similarly veiled by thick aerosol layers.

%


\begin{acknowledgements}
\section*{Acknowledgements}\label{sec:acknowledgements}
Funding for E. Schlawin \edit1{and M. M. M.} is provided by \edit1{the} NASA Goddard Spaceflight Center \edit1{via NASA contract NAS5-02105}.
This research has made use of the SIMBAD database, operated at CDS, Strasbourg, France and NASA's Astrophysics Data System Bibliographic Services.
We respectfully acknowledge the University of Arizona is on the land and territories of Indigenous peoples. Today, Arizona is home to 22 federally recognized tribes, with Tucson being home to the O'odham and the Yaqui. Committed to diversity and inclusion, the University strives to build sustainable relationships with sovereign Native Nations and Indigenous communities through education offerings, partnerships, and community service.
K.O.\ acknowledges funding support from the JSPS KAKENHI Grant Number JP23K19072 and JP21H01141.
T.P.G.\ acknowledges funding support from the NASA Next Generation Space Telescope Flight Investigations program (now JWST) via WBS 411672.07.05.05.03.02 for contributing to the development of the NIRCam instrument.
T.J.B.\ acknowledges funding support from the NASA Next Generation Space Telescope Flight Investigations program (now JWST) via WBS 411672.07.04.01.02.
This work benefited from the 2024 Exoplanet Summer Program in the Other Worlds Laboratory (OWL) at the University of California, Santa Cruz, a program funded by the Heising-Simons Foundation and NASA.
This work is based in part on data collected under the NGTS project at the ESO Paranal Observatory. The NGTS facility is operated by a consortium institutes with support from the UK Science and Technology Facilities Council (STFC) under projects ST/M001962/1, ST/S002642/1 and ST/W003163/1.
\end{acknowledgements}

%

\vspace{5mm}
\facilities{JWST (NIRSpec), JWST (MIRI), HST(WFC3), NGTS, AIT}

The JWST data presented in this article were obtained from the Mikulski Archive for Space Telescopes (MAST) at the Space Telescope Science Institute. The specific observations analyzed can be accessed via \dataset[DOI: 10.17909/tqa7-th94]{https://url.usb.m.mimecastprotect.com/s/5uaECk6WGkCOoN014t2fOCB_Ozz?domain=dx.doi.org}.


\software{astropy \citep{astropy2013,astropy2018v2,astropy2022v5}, 
          \texttt{batman} \citep{kreidberg2015batman},
          \texttt{dynesty} \citep{speagle2020dynesty},
          \texttt{emcee} \citep{foreman-mackey2013emcee},
          \texttt{Eureka!}\ \citep{bell2022eureka},
          \texttt{pysynphot} \citep{lim2015pysynphot},
          \texttt{photutils v0.3} \citep{bradley2016photutilsv0p3},
          \texttt{matplotlib} \citep{Hunter2007matplotlib},
          \texttt{numpy} \citep{vanderWalt2011numpy},
          \texttt{scipy} \citep{virtanen2020scipy},
          \texttt{starry} \citep{luger2019starry},
          \texttt{celerite2} \citep{foreman-mackey2018celerite},
          \texttt{pymc3} \citep{salvatier2016pymc3},
           }



\appendix

\section{Observing Specifications}\label{sec:obsSpecifications}
We observed with the NIRSpec G395H grating along with the F290LP filter in Bright Object Time Series Mode with 16 groups in NRSRAPID mode and a SUB2048 subarray for 1084 integrations on each transit.
We observed two consecutive transits to minimize any possible differences in spot covering fractions of the host star in the program GTO 1185 (observations 18 and 19).
Our exposure durations of 4.62 hours each provided sufficient baseline before and after each transit, which lasted for T$_{14}$ = 0.872 hours.
Previous observations and models of the system with JWST MIRI \citep{kempton2023reflectiveMetalRichGJ1214,gao2023gj1214} indicated very small spectral features, so we stacked two identical observations covering the spectral range from 2.9 to 5.2~\micron\ to increase the signal-to-noise and cover the strongest expected features from CH$_4$ and CO$_2$.
We later combined our NIRSpec broadband light curves with the transit and two eclipse observations as part of the JWST MIRI phase curve to refine the orbital parameters\edit1{, as described} in Section \ref{sec:orbParam}.

\section{Data Analysis}\label{sec:dataAnalysis}

\subsection{tshirt Light Curve Extraction}\label{sec:tshirtExtraction}

We analyzed the \texttt{\_uncal.fits} data products with the Time Series Helper and Integration Reduction Tool (\texttt{tshirt}).
We follow similar steps as previous \texttt{tshirt} analyses of JWST NIRSpec data \citep{ers2023wasp39b_CO2} and NIRCam data \citep{bell2023_methane}.
We use a modified version of the \texttt{jwst} pipeline for stage 1 processing \citep{bushouse2023jwstPipeline1p10p2}.
We skipped the dark current subtraction step as it can add noise.
We used JWST version 1.10.2, Science Data Processing version \texttt{2023\_1a}, CRDS version 11.16.22 and CRDS context \texttt{jwst\_1100.pmap} for both transits and both detectors.

We replaced the reference pixel step with a group-by-group column-by-column 1/f correction.
For this correction, we masked all pixels above 1.5 DN/s as source pixels \edit1{and those that deviate by more than 100 e$^-$ from the median background} and used the remaining pixels for 1/f noise corrections.
We then unfold the group-level image into a time series ordered by pixel and apply a Gaussian process (GP) to interpolate the 1/f noise on to the data pixels. The GP kernel includes 6 Simple-Harmonic Oscillator terms with \texttt{celerite2} \citep{foreman-mackey2018celerite} that mimic a 1/f power spectrum.
Finally, we smooth the GP model with a Savitsky-Golay  polynomial with \texttt{scipy}\citep{virtanen2020scipy} order 3 kernel that is 2001 pixels long.
After the reference pixel step, we proceed with the stage 1 \texttt{jwst} pipeline and set a jump threshold of 10~$\sigma$.
We also apply a temporal cleaning step that examines 150 integrations at a time and replaces pixels that deviate from the median pixel by 20~$\sigma$ by the median value across those 150 integrations.
Finally, we manually divide the image by  \texttt{jwst\_nirspec\_dflat\_0001.fits} for NRS1 and \texttt{jwst\_nirspec\_dflat\_0002.fits} for NRS2.

For the extraction, we \edit1{traced} the source in the images by fitting a Gaussian to the centroid along the Y direction for a single reference integration and then \edit1{fit} a 3 $\sigma$ sigma-clipped 3rd order polynomial to the Gaussian centroids from least squares optimization.
We then fix the source aperture and background aperture to be the same value for all integrations as the reference image.
We use a source aperture full width of 10 pixels, rounded to the nearest pixel (and truncated when at the detector edge).
We use all remaining pixels that are  6 or more pixels away from the centroid for background subtraction by fitting a line with 3 sigma clipping to each column.
We extract the spectrum with a co-variance weighted optimal extraction \citep{schlawin2020jwstNoiseFloorI}, assuming a correlation coefficient of 0.08 between pixels.

\begin{figure}
    \centering
    \includegraphics[width=0.75\linewidth]{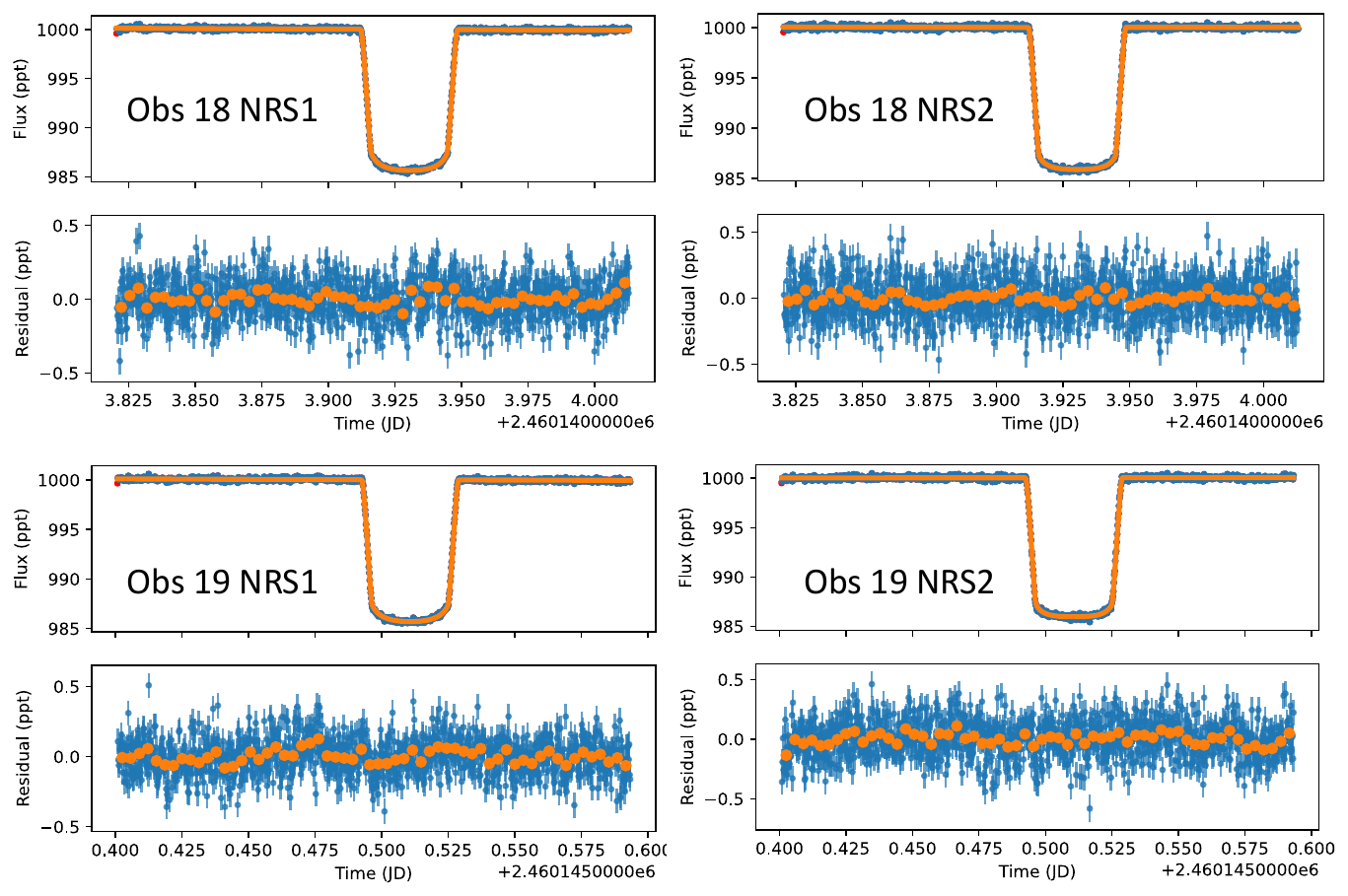}
    \caption{Broadband light curves from the \texttt{tshirt} pipeline separated by detector and observation number.
    The top panels show the light curves at full time resolution and the maximum a posteriori light curve models \citep{luger2019starry}.
    The residuals of the data subtracted by the model are shown in the lower panels, and binned residuals are shown for clarity.
    The first point in the time series (red) is discarded from the fits. }
    \label{fig:BBlc}
\end{figure}

The broadband light curves from each detector show no strong evidence for starspot crossing events so a linear baseline is sufficient to remove the long-term trends, as shown in Figure \ref{fig:BBlc}.
However, examination of the residuals shows that NRS1 and Observation 19 NRS2 exhibited time-correlated noise.
NRS1 generally had higher levels of correlated noise, as in some previous observations with JWST NIRSpec G395H \citep{wallack2024jwstCompass}.
We therefore inflate the error bars in the light curves by simultaneously fitting for the Gaussian sigma in the likelihood function.
In the broadband light curves, the scatter exceeds the theoretical noise from read and photon noise, likely due to residual 1/f noise \citep{schlawin2020jwstNoiseFloorI}, which cannot be perfectly subtracted across the aperture.
In the broadband NRS1 light curves, the standard deviation of the last 309 integrations was 116 (117) ppm for observation 18 (19) compared to the photon and read noise of 100 ppm, and for NRS2 the standard deviation was 135 (145) ppm compared to the photon and read noise of 113 ppm.
For the spectroscopic light curves at R=100, the standard deviation of the residuals had a median that was 1.05 times the photon and read noise limit for NRS1 and 1.15 times the photon and read noise for NRS2.
\edit1{We also re-analyzed the spectra with a divide-by-white approach per detector \citep[e.g.,][]{kreidberg14}, described below in Section \ref{sec:tshirtLCfitting}.}

\subsection{Eureka!\ NIRSpec and MIRI Extraction}\label{sec:eurekaExtraction}
Our second NIRSpec reduction used the open-source Eureka!\ package \citep{bell2022eureka} with version 0.11.dev286+gde5b373b.d20240517, version 1.15.1 of the jwst package, and CRDS version 11.17.26 with CRDS context \texttt{jwst\_1252.pmap}. We also used Eureka!\ to re-reduce the MIRI/LRS full-orbit phase curve observation of GJ 1214 b \citep{kempton2023reflectiveMetalRichGJ1214} using Eureka!\ version 0.10.dev0+g3c109266.d20240308, jwst version 1.13.4, CRDS version 11.17.19, and with CRDS context \texttt{jwst\_1223.pmap}. The Eureka!\ control and parameter files we used are available for download \edit1{(\url{https://doi.org/10.5281/zenodo.13787956}),} and the important parameters are summarized below.

Our NIRSpec Stage 1 analyses largely follow the Stage 1 processing of the jwst pipeline, with some modifications. We increased the jump step rejection threshold from 4.0 to 6.0, used Eureka!'s custom bias correction step, and performed group-level background subtraction to reduce the impact of 1/$f$ noise while masking pixels within 8 pixels of the curved spectral trace. In Stage 2, we skipped the flat-field, photom, and extract\_1d steps. In Stage 3, we masked pixels marked as `DO\_NOT\_USE' in the data quality array, performed a double-iteration background pixel outlier rejection, performed a column-by-column background subtraction for each integration, and then straightened the spectral trace using whole-pixel shifts. We then performed optimal spectral extraction \citep{horne1986optimalE} on the pixels within 5 pixels of the center of the spectral trace, using the median integration to compute the spatial profile. In Stage 4, we then spectrally binned the data and removed any $>$3.5$\sigma$ outliers along the temporal axis compared to a 20-integration wide top hat filter.

Our MIRI/LRS reanalyses turned on the \texttt{firstframe} and \texttt{lastframe} steps in Stage 1 (\edit1{which skip the first and last frames of each integration}) and increased the jump step rejection threshold to 8.0. We did not use the custom 390 Hz MIRI/LRS noise correction step presented in \citet{welbanks2024wasp107b} as we have found that it does not have a significant impact on our planetary spectra. Our Stage 2--4 processing was similar to the NIRSpec processing with slightly different tunings optimized for MIRI/LRS data, including the special background subtraction considerations identified by \citet{bell2024nightsideCloudsDisEqChemWASP43b} and assuming a fixed gain of 3.1 electrons per data number.


\subsection{Stellar Variability Analysis}\label{sec:stellarVariability}
\begin{figure}
    \centering
    \includegraphics[width=0.75\linewidth]{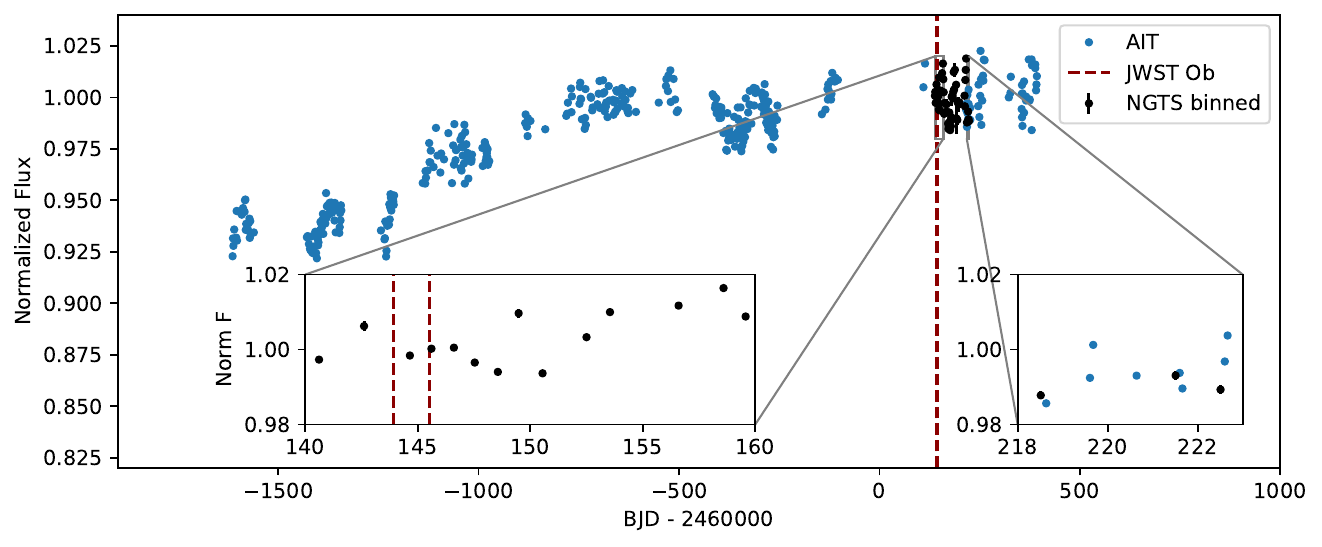}
    \includegraphics[width=0.75\linewidth]{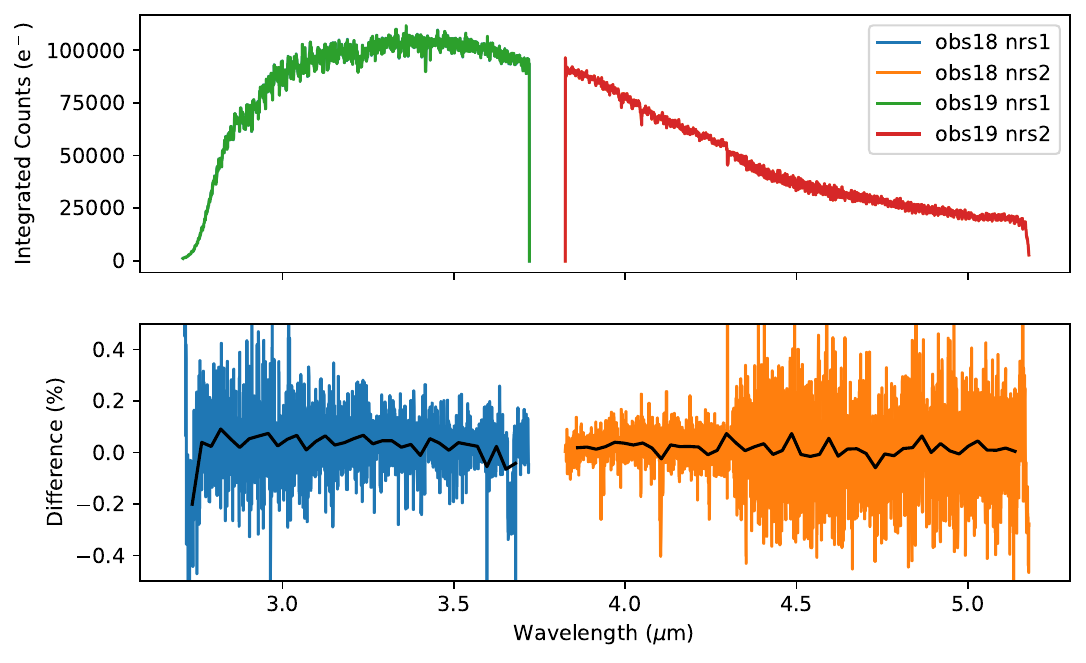}
    \caption{{\it Top:} NGTS and AIT photometry of the GJ 1214 system. A binned version of the NGTS photometry is shown at a cadence of one point per night.
    The epochs of the two JWST NIRSpec observations are highlighted as dark red dashed vertical lines.
    {\it Middle:} The integrated stellar spectra between our two observations over-plotted to check for differences in stellar spot coverage.
    {\it Bottom:} The difference between the two observations, with a median difference of 0.04\% in NRS1 and 0.02\% in NRS2.}
    \label{fig:stellarVariability}
\end{figure}

Previous observations of the GJ 1214 system showed gradual changes in flux of about 3\% over timescales of many months
\citep{mallonn2018rotationPeriodGJ1214bStarspots}.
The stellar rotation period is inferred to be $\gtrsim$50 days with some evidence of a 150 day rotation period \citep{mallonn2018rotationPeriodGJ1214bStarspots,henry2023C14automatedImagingGJ1214}.
We also collected photometry from the Tennesse State University Celestrion 14-inch (C14) Automated Imaging Telescope (AIT) located at Fairborn Observatory \citep{henry2023C14automatedImagingGJ1214} spanning 2006 days.
This uses a Cousins R band filter with 13 comparison stars in the field.
More details on the observing and data reduction are available in \citet{sing2015wasp31bTransmission} and \citet{henry2023C14automatedImagingGJ1214}.
We also collected Next-Generation Transit Survey (\ngts) photometry of GJ 1214 over 84 days including simultaneous coverage of the JWST transit light curve centered on UT 2023-07-20T00:15, enabling us to link long-term photometry from the AIT to the observed transit dates.
\ngts\ \citep{wheatley2018MNRAS.475.4476W} uses an array of twelve 20 cm telescopes at the ESO Paranal Observatory in Chile. 
Each \ngts\ telescope has been designed for high-precision photometry that matches \tess\ for all stars Tmag$>$12 (RMS=400\,ppm in 30\,min), and for stars with Tmag$>$9 by using multiple telescopes \citep[RMS=100\,ppm in 30\,mins; see][]{bryant2020wasp166b}.
Each of the 12 \ngts\ telescopes has a field-of-view of 8 square degrees, providing sufficient reference stars for even the brightest TESS candidates.
We employed 9 of the telescopes for monitoring GJ 1214.
The telescopes observe with a custom filter between 520-890\,nm and are specifically designed for precise photometry of exoplanet transits.
Our photometry is stable night-to-night and is capable of identifying exoplanet transits and stellar variability due to astrophysical sources such as stellar activity \citep[e.g.,][]{ahrer2022WASP39bERS,bryant2021hip41378f}.
We calculate the error in the NGTS light curve by the standard error in the mean per bin where there is one bin per night.
Previous work has indicated that the noise bins as white noise with 1/$\sqrt{N}$ statistics for NGTS, where N is the number of exposures \citep{bryant2020simultaneousTESSNGTStransitW166b,obrien2022scintillationLimitedPhot}.

Figure \ref{fig:stellarVariability} show the long-term photometric monitoring of GJ 1214 with a normalization near the dates of observation.
We assume that the AIT R band and \ngts\ wide filter are both representative of long-term changes on the star, which enables us to link the observed epochs by JWST (vertical dashed lines) to the AIT by normalizing points that are within 0.1 days between AIT and \ngts.
It is clear from the long-term photometric monitoring that our observations occur near the brightest levels from the AIT monitoring, indicating, on average, fewer visible starspots than near BJD-2,460,000 = -1500.
While persistent spots may still exist, our observations indicate relatively less activity on the observable hemisphere of the host star GJ 1214 A than previous epochs that can impact the transmission spectrum of the planet \citep{pont2008hazeHD189733b,rackham2018transitSourceEffect}.

We compared the extracted stellar spectra between our two observations to check for any possible changes in stellar spot coverage on the visible stellar hemisphere.
Our two observations were separated by only one orbital period (1.6 days) and show very little variation in \edit1{average} extracted flux (0.04\% for NRS1 and 0.02\% for NRS2), as shown in Figure \ref{fig:stellarVariability}.
We therefore expect the stellar spot coverage and inhomogeneities to be the same between the two observations.

\subsection{Broadband Orbital Parameters}\label{sec:orbParam}

We obtained new orbital parameters using a broadband light curve fit that included both \edit1{the} NRS1 and NRS2 detectors over both Observation 18 and 19 in Program 1185 as well as the transit and two eclipses from the MIRI LRS phase curve in \edit1{Program} 1803.
The NIRSpec broadband light curves were extracted with \texttt{tshirt}, as described above in Section \ref{sec:tshirtExtraction}, and the MIRI light curves were fit with \texttt{Eureka!} as described above in Section \ref{sec:eurekaExtraction}.
Together, these observations probe three total transits and two eclipses at high cadence and precision. We apply a single orbital solution to fit all light curves simultaneously, and for the transits and the second MIRI eclipse we include a linear systematic baseline for each individual detector and observation. Due to its proximity to the beginning of the observation, the first MIRI eclipse light curve showed visual evidence for a slightly non-linear baseline, so we instead included an exponential baseline for it. We used a single set of quadratic limb darkening parameters per broadband detector (i.e. two parameters for NRS1, two parameters NRS2 and 2 parameters for MIRI LRS). Also, we assume that GJ~1214~b's orbit is tidally circularized and fix the eccentricity to 0 and the argument of periapsis to 90$^{\circ}$. 

We fit for the orbital solution using the Markov Chain Monte Carlo Method with \texttt{emcee} \citep{foreman-mackey2013emcee}, sampling the time of conjunction ($t_0$), orbital period ($P$), semi-major axis ($a/R_\star$), orbital inclination ($cos\left(i\right)$), the individual limb darkening coefficients for each bandpass $j$ ($u_{1,j}$ and $u_{2,j}$), the baseline parameters for each bandpass or observation $k$ (slope $m_k$ and intercept $b_k$ for the linear cases; amplitude $a_k$, exponent $b_k$, and offset $c_k$ for the exponential case), and finally an uncertainty multiplier for each observation ($\sigma_k$). In order to determine a good set of initial values, we first performed a simplified fit with just the transit observations, starting from and using as priors the orbital solution from \citet{cloutier2021gj1214bPreciseMass}. Then, for our final fit including the eclipse observations, we initialize our walkers' positions through random draws from the posterior distributions of this initial fit. We ran this final sampling for 10,000 steps after a 5,000 step burn-in period, which was 15 times the maximum auto-correlation time for any parameter and at least 30 times the typical auto-correlation time and thus sufficient for all parameters to converge. We modeled the transits and eclipses using \texttt{batman} \citep{kreidberg2015batman}. We did not apply any Gaussian priors to any of our parameters. The median and 1-$\sigma$ uncertainties on the best-fit orbital parameters are given in Table \ref{exttab:orbparams}. We then fixed the orbital parameters to these median values when fitting each spectroscopic light curve. 
 

\begin{deluxetable}{cc}[b!]
\tablecaption{Joint orbital parameters from broadband light curve fits. For physical parameters, we use the values from \citep{cloutier2021gj1214bPreciseMass}.}\label{exttab:orbparams}
\tablecolumns{2}
\tablehead{
\colhead{Parameter} &
\colhead{Posterior Value}
}
\startdata
Orbital Period [days] & 1.580 404 341 $\pm$ 0.000 000 079 \\
t0 [BJD$_{TDB}$] & 2 460 143.930 295 03 $\pm$ 0.000 003 58 \\
Inclination, $i$ [deg] & 89.32 $\pm$ 0.03 \\
Scaled Semi-major axis a/$R_*$ & 15.27 $\pm$ 0.02 \\
Eccentricity, $e$ & 0 (fixed) \\
Argument of periastron $\omega$ [deg] & 90 (fixed) \\
\enddata
\tablecomments{Posterior Transit Parameters}
\end{deluxetable}


\subsection{HST Offset Analysis}\label{sec:hstOffset}

As mentioned, GJ 1214b's HST/WFC3 G141 transmission spectrum was previously presented by \citet{kreidberg14}. Rather than fully refit this spectroscopic data for inclusion in our analysis, we instead determined the appropriate vertical offset to apply to the transmission spectrum from \citet{kreidberg14}. Such vertical offsets are known to occur when slightly different orbital parameters are used to fit the spectroscopic light curves, and do not significantly affect the chromatic shape of the spectrum. Therefore, to determine the resulting offset from using the new orbital solution of GJ 1214b that we derive in this work, versus that of \citet{kreidberg14}, we refit the broadband HST/WFC3 G141 light curves presented in \citet{kreidberg14} using our new orbital solution. We refit the broadband light curves from all fifteen observations presented in \citet{kreidberg14} individually, using a \texttt{batman} transit model, a visit-long linear systematic baseline, and an HST orbit-wise exponential systematic ramp as done by \citet{kreidberg14} and others. The only free parameters in this fit were the planet-star radius ratio $R_p/R_\star$ and the coefficients for these systematic models. \citet{kreidberg14} discarded the 13th and 14th visits due to potential starspot crossings, but we do not see evidence of this affecting our results and kept them in our analysis. We did discard the 10th visit as this observation had a Fine Guiding Sensor failure \citep{kreidberg14}, and as a result the best-fit transit depth was significantly errant from the other visits. The best-fit transit depths from the other 14 visits were all consistent within 2.8$\sigma$ and scattered with a standard deviation of $\sim$118~ppm, and we calculated the mean transit depth to be 13,370 $\pm$ 30~ppm. Comparing this to the mean broadband transit depth of 13,489 ppm computed by \citet{kreidberg14} with the \texttt{model-ramp} value, we derive a 119~ppm offset due to our new orbital solution. We therefore apply this offset to the transmission spectrum from \citet{kreidberg14}. 

\subsection{tshirt Light Curve Fitting}\label{sec:tshirtLCfitting}

We proceeded to fit the light curves at a constant resolving power of R=100.
We fixed orbital parameters with the mean Posterior values from Table \ref{exttab:orbparams}.
The light curves are fit in two iterations with different limb darkening treatments.
In the first iteration, we used uninformative free quadratic limb darkening coefficients $q_1$ and $q_2$ \citep{kipping2013uninfomativePriorsQuad}.
In the second iteration, we fixed the $q_1$ and $q_2$ parameters.
After the first iteration, we compared these coefficients to a Phoenix model using \texttt{exotic-LD} \citep{grant2024exoticLDjoss,husser2013phoenix}, and derived a scaling factor to the Phoenix limb darkening model.
We used \texttt{pymc3} and No-U Turns sampling \citep{salvatier2016pymc3} to fit the light curves as in \citet{schlawin2024wasp69bpublished,bell2023_methane}.
We assume a linear baseline trend with time and an astrophysical model with \texttt{starry} \citep{luger2019starry}.
We clipped any points that deviated from the fit by more than 6~$\sigma$ and simultaneously fit for the light curve error to allow for increased noise over the photon and read noise limit.

\begin{figure}
    \centering
    \includegraphics[width=0.49\linewidth]{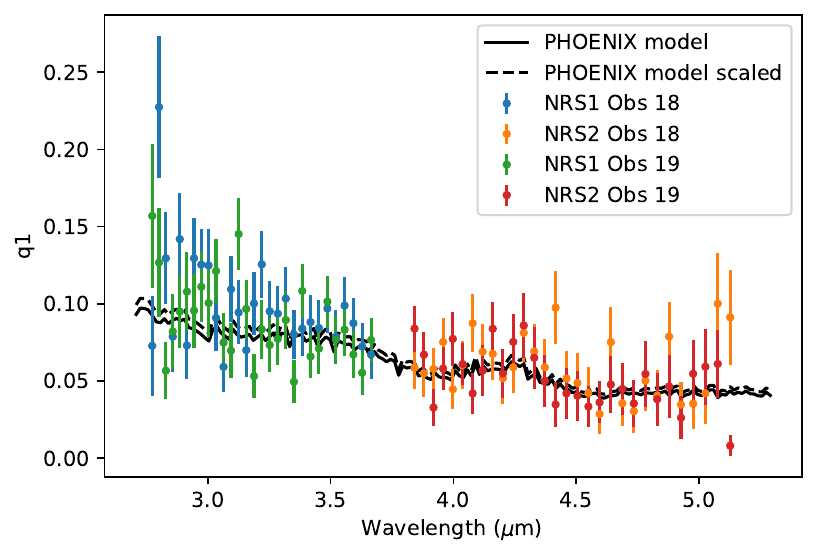}
     \includegraphics[width=0.49\linewidth]{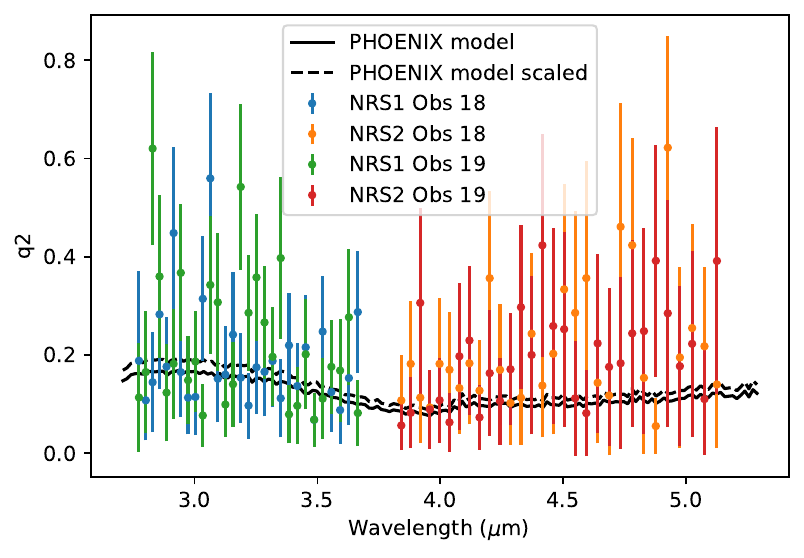}
    \caption{The calculated limb darkening coefficients are very close to the Phoenix stellar model.
    We fit for a scale factor for this model and then fix the q$_1$ and q$_2$ coefficients for a second iteration of spectroscopic light curve fitting.
    Stellar parameters are T$_{eff}$=3300 K, [M/H]=0.5, log(g)=5.0}
    \label{fig:LDcoeff}
\end{figure}

Figure \ref{fig:LDcoeff} shows the fitted $q_1$ and $q_2$ limb darkening coefficients as compared to a Phoenix model for those same coefficients.
We then perform a linear least squares fit with a scale factor of 1.066 $\pm$ 0.001 for $q_1$ and 1.145 $\pm$ 0.004.
In the second iteration, we fixed the quadratic limb darkening coefficients to those from the scaled Phoenix model.
Other free parameters in the fit were a linear slope in time and re-scaling factor.
We also checked for timing offsets between detectors \citep{wallack2024jwstCompass} or with wavelength and find that the timing is consistent with the median values in our ephemeris from Table \ref{exttab:orbparams} to within $\pm$5 seconds or smaller.
We also did another spectroscopic analysis with a divide-by-white approach \citep[e.g.,][]{kreidberg14}, which eliminates common-mode noise; we found that the spectra were consistent within 5 ppm, and the transit depth errors were within 8\% of each other.

\subsection{Eureka!\ Light Curve Fitting}\label{sec:eurekaLCfitting}

For the NIRSpec data, our astrophysical model consisted of a batman transit function \citep{kreidberg2015batman}, with limb-darkening coefficients fixed to a PHOENIX model limb-darkening spectrum computed using the ExoTiC-LD package \citep{grant2024exoticLDjoss,husser2013phoenix}. For the MIRI/LRS wavelength region, we needed to linearly extrapolate the PHOENIX limb-darkening model for our batman transit model. Similar to the tshirt methods applied above, we also tried freely fitting for the uninformative quadratic limb darkening coefficients $q_1$ and $q_2$ \citep{kipping2013uninfomativePriorsQuad}, and we found that our fitted coefficients were generally consistent with the PHOENIX model limb-darkening coefficients. Our MIRI/LRS astrophysical model also included a batman eclipse model (accounting for the differences in light-travel time between transit and eclipse) and a second-order sinusoidal phase variation model (with all phases forced to have flux greater than or equal to 0 ppm).

The systematic noise model for NIRSpec consisted of a linear trend in time, a linear de-correlation against the trace's spatial position and PSF width, a GP with a Mat\'ern-3/2 kernel (as implemented by \texttt{celerite2}; \citealt{celerite1, celerite2}) as a function of time to account for any residual red noise, and a white noise multiplier to account for any additional white noise in the light curves. Our MIRI/LRS systematic noise model included all of the same model components as our NIRSpec model in addition to a single exponential ramp, following the recommendations of \citet{bell2024nightsideCloudsDisEqChemWASP43b}. We removed the first 700 integrations from the MIRI/LRS data to reduce the correlation between the initial systematic exponential ramp and the planetary orbital phase variations, as recommended by \citet{bell2024nightsideCloudsDisEqChemWASP43b}. We also discarded integrations 4220--4330, 8540--8650, 12860--12970, 17180--17290, and the last 100 integrations that were affected by high-gain antenna moves in the MIRI/LRS phase curve. Because the detector settling time for NIRSpec is much faster, we only removed the first 20 integrations from each of the NIRSpec observations. We then separately modelled each light curve, using the dynesty nested sampling package \citep{speagle2020dynesty} with `multi' bounds, `rwalk' sampling, a convergence criterion of $d\log\mathcal{Z}<0.01$, and 121 live points for the simpler NIRSpec transit light curves and 256 live points for the MIRI/LRS phase curve observations.

Our white noise levels in the NIRSpec spectrum were between 1.25 and 1.40 times the stellar-photon limited noise level for both detectors and both visits. A visual examination of our Allan variance plots \citep{Allan1966} shows that our included GP has accounted for any otherwise-unmodelled red noise, although the amplitude of the GP mean model was generally $\lesssim$20 ppm with the exception of the first channel of NRS1 from observation 18 and the last channel of NRS2 from observation 19, which had amplitudes of $\sim$200 ppm. Finally, we also compared the transit depths for each spectral bin between our two NIRSpec visits and found an error-weighted offset of $-28.6 \pm 9.6$ ppm of the second NIRSpec visit with respect to the first visit. After moving the second visit upward to remove that offset, we find that $\sim$81\% of points agree between the two visits within 1 sigma, while the two most discrepant points still only disagree at the 2--3$\sigma$ level.

For MIRI/LRS, our white noise levels were between 1.10 and 1.15 times the stellar-photon limited noise level below 10.5 $\mu$m, beyond which the noise steadily increased to 1.28 times the stellar-photon noise limit. Again, our Allan variance plots \citep{Allan1966} show that our GP accounted for any otherwise-unmodelled red noise. The GP for the MIRI/LRS data had a larger amplitude compared to our NIRSpec fits ($\sim$200 ppm) with varying timescales between different channels, although the GP mean model seemed to have the lowest amplitude and/or longest timescale between 8 and 10 $\mu$m. Our re-reduced MIRI/LRS transmission spectrum was 29 ppm lower on median than that of \citet{kempton2023reflectiveMetalRichGJ1214}, in large part due to the changes in orbital parameters, and our transit depth uncertainties were 40\% larger on median. Our re-reduced MIRI/LRS eclipse spectrum was consistent with that of \citet{kempton2023reflectiveMetalRichGJ1214} within 1$\sigma$ at all points, differing by only -8 ppm on median and with 17\% larger uncertainties on median. Our nightside emission spectrum was also consistent with that of \citet{kempton2023reflectiveMetalRichGJ1214} within 1$\sigma$ at all points and had uncertainties that were $\sim$37\% larger than those of \citet{kempton2023reflectiveMetalRichGJ1214} on median.

Finally, we compared the integrated counts averaged over Observation 18 and compared this to the integrated counts averaged over Observation 19.
The pseudo-continuum levels of the star are consistent to better than 0.1\%, which indicates that there was no significant change in stellar spot or plage coverage between the two successive transits on 2023-07-18 and 2023-07-19.
This is the same inference from the  \texttt{tshirt} analysis shown in Figure \ref{fig:stellarVariability}.

\section{Atmospheric Models}\label{sec:method_model}
We interpret the observed transmission spectrum by comparing the data with atmospheric radiative transfer models.
To ensure the robustness of the inference for atmospheric compositions, we utilized two types of atmospheric models, one assuming \edit1{chemical} equilibrium and a chemically-agnostic free retreival.
For all models, we assume a stellar radius of $0.215~{\rm R_{\rm sun}}$ and a planetary mass of $8.17~{\rm M_{\rm Earth}}$ \citep{cloutier2021gj1214bPreciseMass}.
We note these are consistent with the newer values derived from JWST transit light curve fitting for the system of $0.2162^{+0.0025}_{-0.0024} ~{\rm R_{\rm sun}}$ for the host star and $8.41^{+0.36}_{-0.35}~{\rm M_{\rm Earth}}$ for planet b \citep{Mahajan+24_gj1214update}.

\subsection{Free Retrieval}\label{sec:freeRetrieval}

We provide some additional details about the Aurora model used to fit the spectrum, described in Section \ref{sec:results}.
Aurora solves radiative transfer for a parallel-plane atmosphere in transmission geometry assuming hydrostatic equilibrium. The vertical temperature structure of the atmosphere is parameterized using the six parameter prescription from \citet{Madhusudhan2009}.
Given previous suggestions of a cloudy atmosphere for GJ1214b, we consider the possibility of inhomogeneous clouds and hazes using the 2 sector prescription from \citet{Welbanks&Madhusudhan21Aurora} which considers the linear combination of a cloud-free and cloudy and hazy atmosphere \citep[e.g.,][]{line2016nonUclouds}. Bayesian inference is performed using \texttt{PyMultinest} \citep{Buchner+14_pymultinest, feroz2009multinest}.
The sources of opacity are obtained from HITRAN \citep[][]{Rothman2010, Richard2012} and ExoMol \citep[][]{Tennyson2016} as described in \citet[][]{Welbanks&Madhusudhan21Aurora, welbanks2024wasp107b}.

In total, our atmospheric retrieval has 29 parameters: 14 chemical abundances, 6 for the pressure-temperature profile, 4 for inhomogeneous clouds and hazes, 1 for the reference pressure, 1 for the planet radius at the reference pressure, and 3 for offsets between NIRSpec G395H NRS2 and HST-WFC3, NIRSpec G395H NRS1, and MIRI-LRS. 
The inferred atmospheric properties of GJ1214b are suggestive of signatures of CH$_4$ with an abundance of $\log_{10}\text{X}_{\text{CH}_4}=-2.25 ^{+ 0.72 }_{- 1.08 }$ in a CO$_2$-rich atmosphere making up $91^{+ 8}_{-45}$\% of the bulk atmospheric composition.
The abundances of all other gases remain largely unconstrained and so are the cloud properties, only suggesting a generally cloudy atmosphere with a 2$\sigma$ lower limit of 25\% cloud and haze cover. The retrieved temperature near the photosphere at 100mbar is consistent with the equilibrium temperature of the planet with a retrieved value of T$_{100\text{mbar}}=556^{+192}_{-103}$~K. 

\subsection{Chemically-Consistent Retrieval}\label{sec:chemConsistentRetrieval}
We perform the chemically-consistent retrieval using an open source radiative transfer tool \texttt{CHIMERA} \citep{line2013chimera,Mai&Line19cloudAssumptions}.
For the atmospheric temperature-pressure (TP) profile, we adopted the analytical model of \citet{guillot2010radEquilibrium} to flexibly vary the possible TP profiles, where we vary the heat redistribution factor, the infrared opacity and the visible-to-infrared opacity ratio as free parameters to be retrieved.
The planetary Bond albedo is fixed to $A_{\rm b}=0.5$ as constrained by the MIRI phase curve \citep{kempton2023reflectiveMetalRichGJ1214}.
We then use the open-source equilibrium chemistry code \texttt{FastChem2} \citep{stock2022fastchem2} to calculate the molecular abundances in thermochemical equilibrium in each vertical layer on the fly.
To account for possible disequilibrium effects, we introduce a quench pressure following \citep{morley2017gj436bRetrieval}.
We express aerosol opacity by a vertically constant opacity with a power-law wavelength dependence, $\kappa_{\rm aer}=\kappa_{\rm 0}(\lambda/\lambda_{\rm 0})^{-\alpha}$. 
This parameterization with variable $\alpha$ can flexibly explore the possible aerosol candidates proposed by previous microphysical studies, such as KCl clouds composed of spherical/aggregate particles \citep{Ohno&Okuzumi18microphysicalCloudModelsGJ214gj436b,gao2018microphysicsGJ1214clouds,Ohno+20fluffyaggregate} and photochemical hazes composed of spherical/aggregate particles \citep{Adams+19aggregateHazes,Lavvas+19photochemicalHazes,gao2023gj1214}. 
We then use \texttt{PyMultinest} \citep{Buchner+14_pymultinest}, a Python implementation of \texttt{Multinest} package \citep{feroz2009multinest}, to obtain the posterior distributions for atmospheric metallicity, C/O ratio, quench pressure $P_{\rm quench}$, atmospheric infrared opacity $\kappa_{\rm IR}$, visible-to-infrared opacity ratio $\gamma$, aerosol opacity $\kappa_{\rm Aer}$ at $0.43~{\rm {\mu}m}$, power-law index of wavelength dependence of aerosol opacity $\alpha$, planetary reference radius at $10~{\rm bar}$, and instrumental offset of NRS2, HST, and MIRI data as compared to NRS1 bandpass.
We used the nested sampling live points of $N_{\rm LP}=500$.
A Gaussian prior with a standard deviation of $38~{\rm ppm}$ is adopted for the instrumental offset parameter, while we adopted uniform priors for the remaining parameters.

\subsection{Bayesian Model Comparison Between a Flat Line and CO$_2 +$ CH$_4$-containing Atmosphere}\label{sec:bayesFlatLine}

\begin{figure*}
    \centering
    \includegraphics[width=\linewidth]{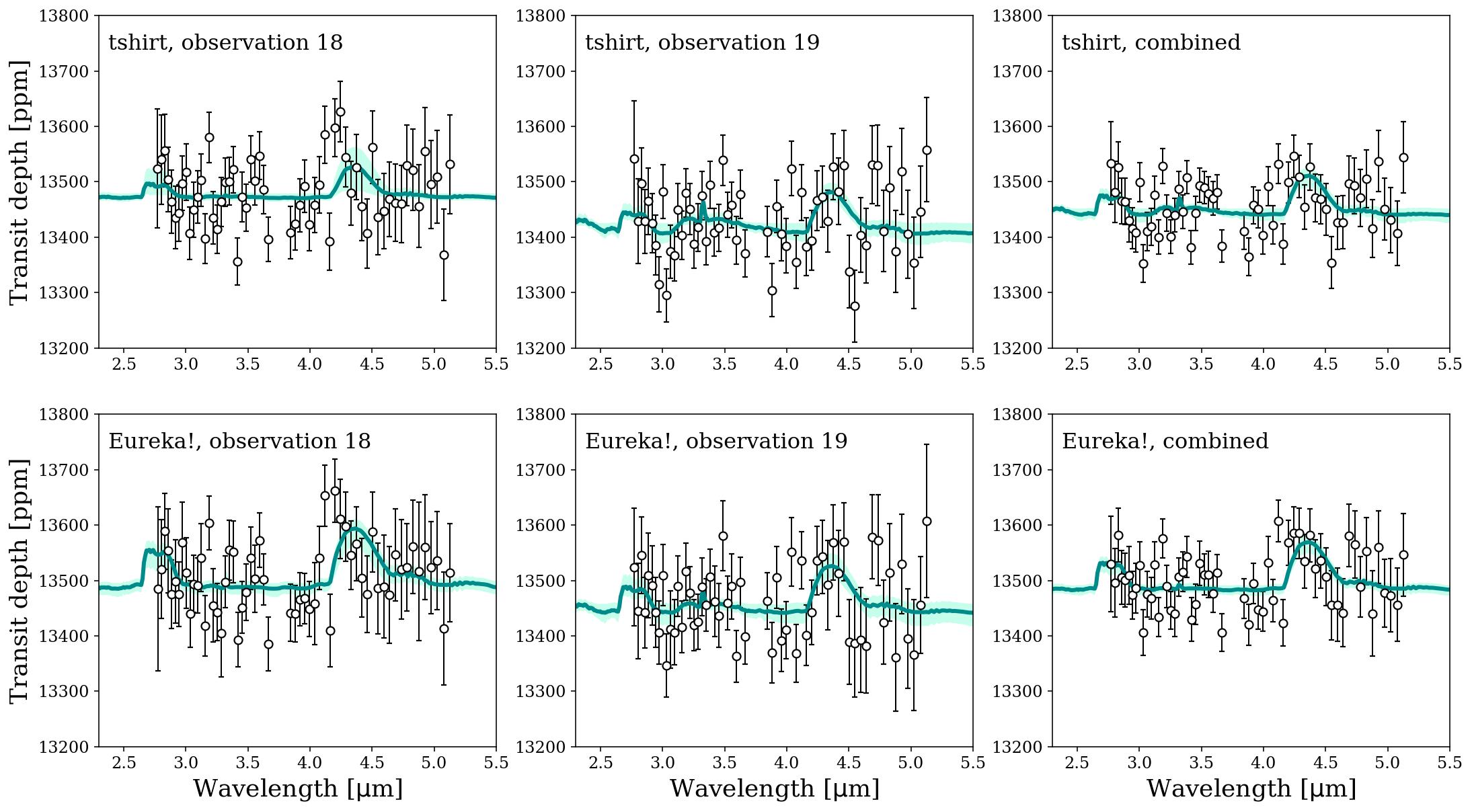}
    \caption{Median CO$_2$+CH$_4$ atmospheric models for spectrum taken at each individual observational epoch. The upper and lower row shows the modal-data comparison for \texttt{tshirt} and \texttt{Eureka!} data reduction pipeline, respectively.}
    \label{fig:test_CO2CH4}
\end{figure*}

\begin{table*}[t]
  \caption{Summary of the model comparison between flat line and CO2 atmosphere models}\label{table:test1}
  \centering
  \begin{tabular}{l || c| c } \hline
     & Flat line & CO$_2$+CH$_4$ atmosphere \\ \hline \hline
    tshirt (obs18)  & $\chi^2_{\rm \nu}=1.28$ & $\chi^2_{\rm \nu}=1.25$ / $\ln{B}=1.18$ ($2.13\sigma$) \\
    tshirt (obs19)  & $\chi^2_{\rm \nu}=1.40$ & $\chi^2_{\rm \nu}=1.30$ / $\ln{B}=2.73$ ($2.84\sigma$)\\   
    tshirt (combined) & $\chi^2_{\rm \nu}=1.76$ & $\chi^2_{\rm \nu}=1.62$ / $\ln{B}=3.96$ ($3.28\sigma$) \\   
    Eureka! (obs18) & $\chi^2_{\rm \nu}=1.11$ & $\chi^2_{\rm \nu}=1.00$ / $\ln{B}=3.31$ ($3.06\sigma$) \\
    Eureka! (obs19) & $\chi^2_{\rm \nu}=1.06$ & $\chi^2_{\rm \nu}=0.94$ / $\ln{B}=3.05$ ($2.97\sigma$) \\    
    Eureka! (combined) & $\chi^2_{\rm \nu}=1.43$ & $\chi^2_{\rm \nu}=1.21$ / $\ln{B}=5.51$ ($3.75\sigma$)
  \end{tabular}
\end{table*}

We \edit1{compared} the goodness-of-fit between the flat line and simplified atmospheric models. 
For the flat line model, we fit the observed spectrum with a wavelength-independent transit depth, and introduce a potential offset between the two detectors, NRS1 and NRS2, which was shown to matter for planets with weak spectral features \citep{alderson2024compass}.
For the atmospheric model, we used \texttt{CHIMERA} to compute the transmission spectrum for an atmosphere made up of only CO$_2$ and CH$_4$, since previous studies predicted potentially observable features of CO$_2$ and CH$_4$ at the NIRSpec bandpass even if the spectrum is featureless at the HST and MIRI bandpass \citep{gao2018microphysicsGJ1214clouds,Lavvas+19photochemicalHazes,Ohno+20fluffyaggregate,gao2023gj1214}.
As the simplest atmospheric model that is capable of producing CO$_2$ and CH$_4$ features, we assumed an isothermal atmosphere of $T=500~{\rm K}$, which is the intermediate temperature between the dayside and nightside temperatures retrieved from the MIRI emission spectrum \citep{kempton2023reflectiveMetalRichGJ1214} and introduced a gray cloud opacity as another parameter.
In summary, the flat line model has 2 free parameters (planetary radius, offset parameter), while the atmospheric model has 4 free parameters (planetary reference radius, offset parameter, CH$_4$-to-CO$_2$ abundance ratio, cloud opacity). 
We used \texttt{PyMultinest} \citep{Buchner+14_pymultinest} to obtain the posterior distributions and Bayesian evidence for each flat line and CO$_2$+CH$_4$ model at each observation 18, 19, and the combined data.
We used \edit1{200 nested sampling live points} for both models.

Table \ref{table:test1} summarizes the reduced chi-squared values $\chi^2_{\rm \nu}$ and logarithmic Bayes factor of the CO$_2$+CH$_4$ model against the flat line model for each observational epoch and two independent data reduction pipelines.
In terms of $\chi^2_{\rm \nu}$, CO$_2$+CH$_4$ models are preferred over the flat line at every observation and data reduction configuration.
The Bayes factor \citep{Trotta08bayesianInference} indicates weak ($\ln{B}>1.0$) to moderate ($\ln{B}>2.5$) evidence of the CO$_2$+CH$_4$ model for each observation epoch.
The combined data \edit1{from} two observational epochs yielded moderate to strong ($\ln{B}>5.0$) evidence \edit1{for the} CO$_2$+CH$_4$ model.
In summary, our observed data prefers the CO$_2$+CH$_4$ model over a flat line by $2$--$3\sigma$ confidence for each individual observation and $3$--$4\sigma$ for the combined data, where the $\sigma$ value is converted from the Bayes factor following the method of \citep{Welbanks&Madhusudhan21Aurora}.
We stress that, as demonstrated in Figure \ref{fig:test_CO2CH4}, our observed spectrum consistently shows bumps in \edit1{${\sim}4.3$~\micron\ and ${\sim}2.8$\micron} that are consistent with the CO$_2$ features regardless of the observation epoch and the data reduction pipelines.
This also highlights the repeatability of our inferences across the two observations.

\subsection{Stellar spot contamination}\label{sec:spotContamination}
\begin{figure}
    \centering
    \includegraphics[width=\linewidth]{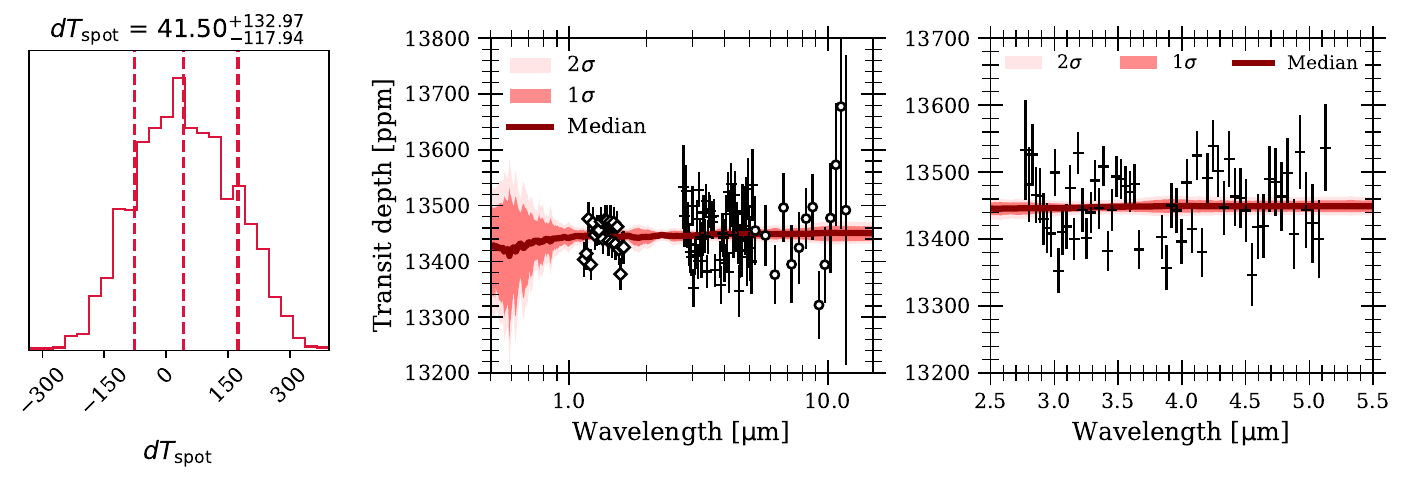}
    \caption{A posterior distribution of the temperature difference in stellar spot from mean effective temperature.
    The left panel shows the spot contrast temperature (in K).
    The middle and right panels show the retrieved spot contamination spectrum.}
    \label{fig:retrievalPlot3}
\end{figure}

We evaluated whether \edit1{the} observed spectrum may appear to have features because of stellar spot contamination that mimics molecular features \citep{moran2023gj486bwater}, even if the actual atmospheric spectrum is flat.
For computing the spot contamination spectrum, we adopt the prescription for spot contamination proposed elsewhere \citep{Iyer&Line20} assuming the stellar photospheric properties of $T_{\rm eff}=3250~{\rm K}$, [Fe/H]$=0.29$, log$(g)=5.026$ \citep{cloutier2021gj1214bPreciseMass}.
For the spot properties, we fix the spot coverage to be $f_{\rm s}=0.032$ following \citep{Rackham17_spot} and use \texttt{PyMultinest} to obtain the posterior distribution of the temperature deviation from the mean effective temperature $dT_{\rm s}$ in addition to the instrumental offset parameters.
We adopt a Gaussian prior centered on $dT_{\rm s}=0$ with a standard deviation \edit1{of} 10\% of \edit1{the} mean effective temperature ($325~{\rm K}$) following \citep{Rackham17_spot}. 
The stellar spectrum is interpolated from the SPHINX grid \citep{Iyer+23sphinx} that is a dedicated stellar grid for cool M dwarfs.
We assumed that the atmospheric spectrum is completely flat and investigated how well spot contamination alone could explain the observed spectrum.

Figure \ref{fig:retrievalPlot3} shows the posterior distribution of the difference in the spot temperature from the mean effective temperature.
The retrieved spot temperature difference is $dT_{\rm spot}=41.5^{+132}_{-117}~{\rm K}$.
This temperature contrast does affect the spectrum at visible wavelengths as suggested by \citep{Rackham17_spot} but barely influences the spectrum at the NIRSpec bandpass.
In fact, the retrieved spot spectrum is nearly the same as a flat featureless spectrum because there is no H$_2$O feature in our measured GJ 1214 b spectrum.

\subsection{Panchromatic Spectrum}\label{sec:panchromatic}

\begin{figure}
    \centering
    \includegraphics[width=0.80\linewidth]{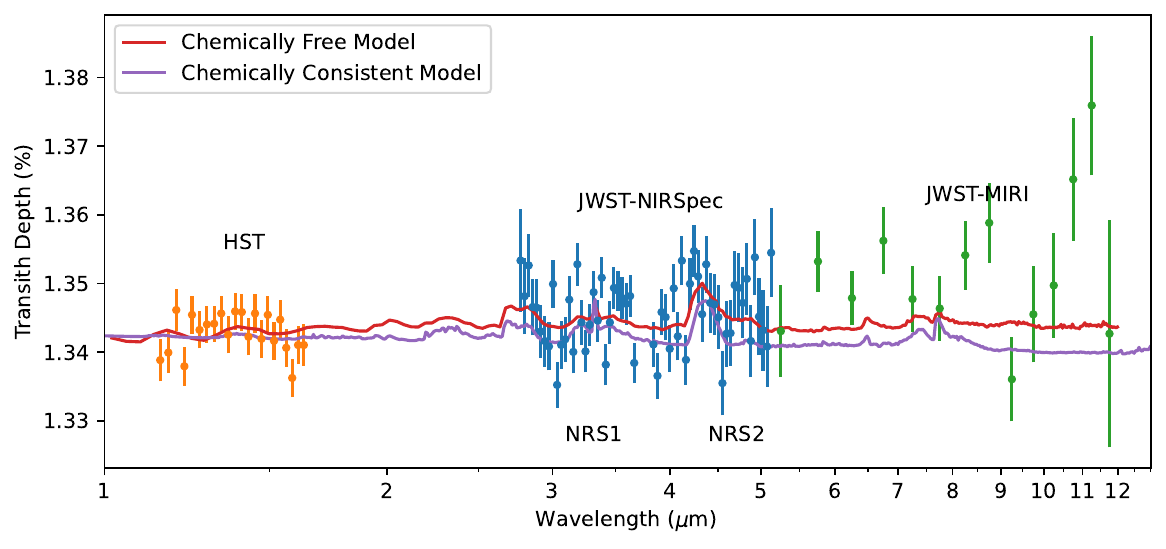}
    \caption{\textbf{GJ 1214 b's Panchromatic Transmission Spectrum.}:
    We plot the combined results from HST WFC3 \citep{kreidberg14}, JWST NIRspec (this work) and JWST MIRI \citep{kempton2023reflectiveMetalRichGJ1214}, re-analyzed with our best-fit orbital parameters.
    The data are offset according to the best-fit chemically-free model, as described in Section \ref{sec:panchromatic}.
    While our JWST NIRSpec spectrum is best fit with features of CO$_2$ and (weakly) CH$_4$, the models predict virtually featureless spectra in the HST and JWST MIRI wavelengths, consistent with previous studies \citep{kreidberg14,gao2023gj1214}.
    Future high precision characterization in the 2~\micron\ to 5~\micron\ wavelength range will lead to better characterization of the carbon, oxygen and sulfur content of the atmosphere and allow for comparisons to planet formation models.
    }
    \label{fig:panchromaticSpec}
\end{figure}

We combined together the HST WFC3, JWST NIRSpec and JWST MIRI LRS transmission spectra to better constrain the spectrum over a wide wavelength range using common orbital parameters between datasets.
We combine all three of the these data sets with the \citet{kreidberg14} HST WFC3 spectrum, shifted as described in Section \ref{sec:hstOffset}, the \texttt{tshirt} version of JWST NIRSpec described in Section \ref{sec:tshirtLCfitting} and the \texttt{Eureka!} MIRI LRS analysis described in Section \ref{sec:eurekaLCfitting}.
We fit simultaneously for an offset between HST WFC3, NIRSpec NRS1 and JWST MIRI LRS all relative to NIRSpec NRS2.
We found a best fit with HST WFC3 shifting up by 56 ppm, JWST NIRSpec NRS1 shifting down by 2 ppm and JWST MIRI LRS shifting up by 50 ppm all relative to JWST NIRSpec NRS2 and plot the resulting points with error bars in Figure \ref{fig:panchromaticSpec}, 

Our two atmospheric models can fit the \cotwo\ and \methane\ features in the NIRSpec bandpass and simultaneously fit the featureless regions observed in HST WFC3 (1.1 to 1.6~\micron) and JWST MIRI LRS (5 to 12~\micron).
Both the chemically-free best-fit model described in Sections \ref{sec:results} and \ref{sec:freeRetrieval} as well as the chemically-consistent best-fit model described in Sections \ref{sec:results} and \ref{sec:chemConsistentRetrieval} are plotted in Figure \ref{fig:panchromaticSpec}.
These models predict that future high precision observations with JWST likely will benefit from targeting the 2.0~\micron\ to 5~\micron\ wavelength region to better constrain the molecules in GJ 1214 b.
Future models of the planet with atmospheric hazes may reveal whether organic features at 6~\micron\ and 3.4~\micron\ \citep{Corrales+23photochemHazesCtoO,He+23organichazesWaterRich} are visible in GJ 1214 b. 
We will further discuss the interpretation of the panchromatic spectrum based on atmospheric forward models in our companion paper (Ohno et al. in prep).

\section{Additional Checks for Significance}\label{sec:additionalSigChecks}
\subsection{Methane Differential Lightcurves}\label{sec:featureLC}
\begin{figure}
    \centering
     \includegraphics[width=0.49\linewidth]{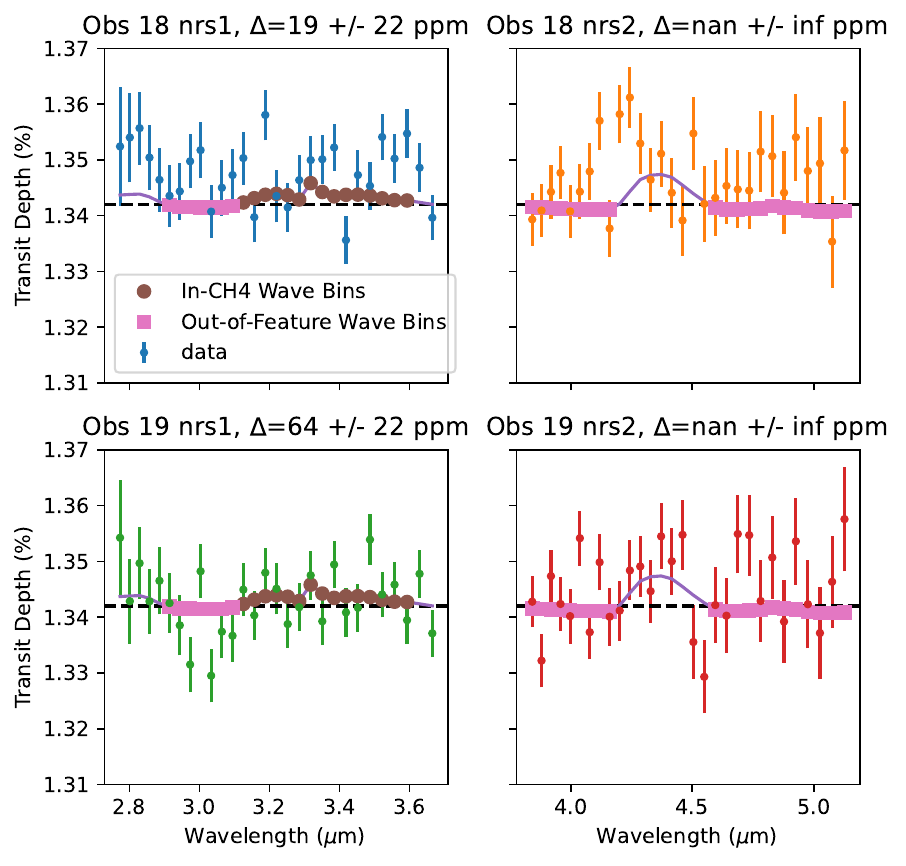}
    \includegraphics[width=0.49\linewidth]{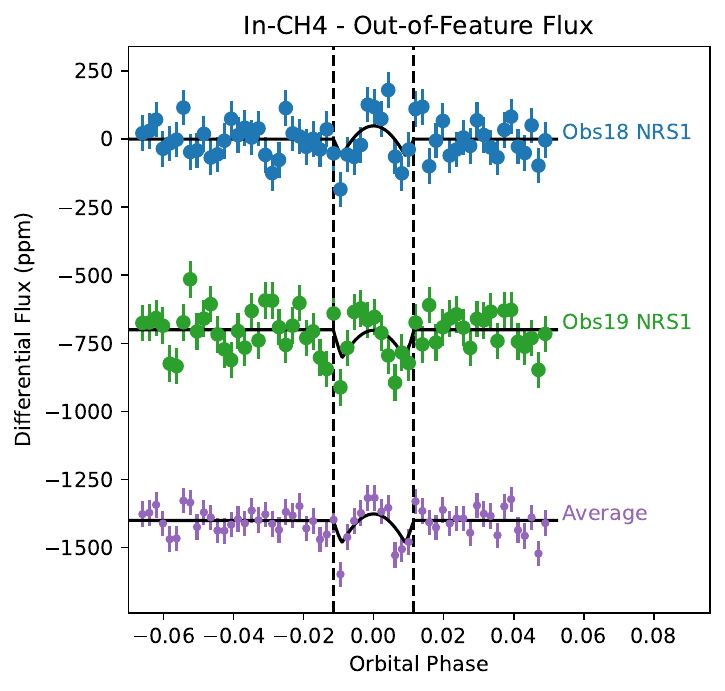}
    \caption{{\it Left:} Wavelength selections for CH$_4$ and Out-of-Feature points.
    {\it Right:} Differential light curve inside and outside of significant CH$_4$ features. Colors and descriptions are the same as for Figure \ref{fig:CO2lc}.}
    \label{fig:CH4lc}
\end{figure}

\edit1{We also repeated the differential lightcurve analysis} from Figure \ref{fig:CO2lc} for the \methane\ feature at 3.3~\micron, \edit1{and show the result} in Figure \ref{fig:CH4lc}, which shows less clear evidence of absorption than \cotwo, and only is possible on the NRS1 detector.
\edit1{The differential lightcurve is dominated by limb darkening differences and the 3.0~\micron\ transit depth dip in Observation 19.}

\subsection{Gaussian Feature Test}\label{sec:gaussFeature}

\edit1{
As visible in Figure \ref{fig:CO2lc}, observation 18 exhibited  a $\sim$ 200 ppm increased transit depth around 4.25~\micron\ that is shifted blue-ward from the expected peak opacity of \cotwo.
It is possible that systematic or random errors contribute to this peak and since it overlaps with \cotwo, this peak may mimic the detection of \cotwo\ in the observation-averaged spectrum in Figure \ref{fig:differentAnalyses}.}

\edit1{We consider two models for the planet spectrum to help assess whether an anomalous spike (modeled by a Gaussian) may fit the spectrum better than \cotwo.
We first fit the observation-averaged spectrum with a Gaussian feature (denoted as ``Free Gaussian'') and then we fit it again with Gaussian that fixed in width and location to the known 4.3~\micron\ opacity feature (denoted as ``\cotwo\ Gaussian'').
We use \texttt{dynesty 2.1.4} \citep{speagle2020dynesty} to sample the models with dynamic nested sampling \cite{higson2019dynNestingSampling} and the bounding method from \citet{feroz2009multinest}.
The Free Gaussian model includes a baseline transit depth in NRS1, a baseline transit depth in NRS2 and 3 free parameters for the Gaussian: its amplitude (A$_F$), centroid ($\mu_\mathrm{F}$) and width ($\sigma_\mathrm{F}$).
We assume Normal priors on each detector's mean transit depths of 13,450$\pm$500 ppm, a Gaussian amplitude of 0$\pm$200 ppm.
We assume a truncated Normal prior and a Gaussian width of 0.1$\pm 0.05$~\micron\ with limits from 0.05~\micron\ to 3~\micron, with the lower limit set to avoid single points.
We assume a Uniform prior on the Gaussian centroid ($\mu_\mathrm{F}$) from 4 to 5~\micron.
The \cotwo\ Gaussian model includes the same baseline transit parameters and the same amplitude parameter (A$_{\mathrm{CO}_2}$) as the Free Gaussian model, along with the same priors.
However, the \cotwo\ Guassian model is fixed in centroid ($\mu_{\mathrm{CO}_2}$) and width 
($\sigma_{\mathrm{CO}_2}$)
to the best-fit value from the chemically-consistent model in Section \ref{sec:chemConsistentRetrieval}.}

\begin{figure*}
    \centering
     \includegraphics[width=0.99\linewidth]{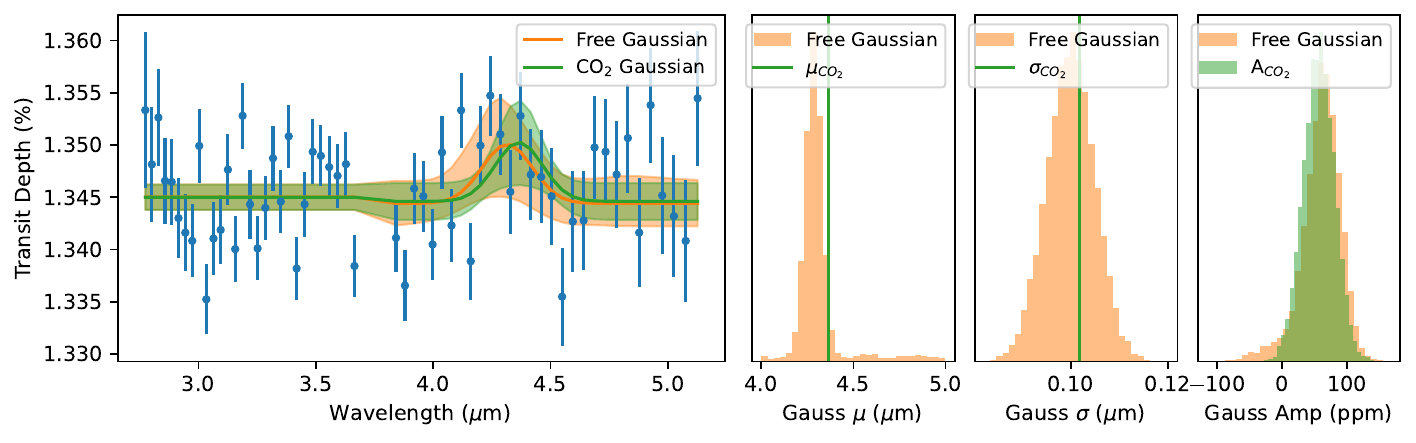}
    \caption{{\it Left:} A Free Gaussian fit to the spectrum as well as a Gaussian that has the same centroid and width as expected from CO$_2$.
    {\it Middle and Right:} Posterior distributions of the Free Gaussian centroid and width as compared to the best-fit values Gaussian for a CO$_2$-rich model.}
    \label{fig:GaussianFeature}
\end{figure*}

\edit1{Figure \ref{fig:GaussianFeature} shows the posterior distributions of the Free Gaussian and the \cotwo\ Gaussian models.
We find that the Free Gaussian fit and the \cotwo\ Gaussian have similar amplitudes: A=$60^{+26}_{-30}$ vs A$_{\rm{CO}_2}$ = $62^{+23}_{-23}$ ppm and widths $\sigma$=$0.104^{+0.036}_{-0.035} $~\micron\ vs $\sigma_{\rm{CO}_2}$=0.102~\micron.
The Gaussian feature is shifted as compared to \cotwo: $\mu$=4.288$^{+0.064}_{-0.047}$ vs $\mu_{\mathrm{CO}_2}$=4.365, as expected from the blueward peak in Observation 18 (Figure \ref{fig:CO2lc}).
However, the shift is not statistically significantly preferred (equivalent to a 1.2~$\sigma$ event for a Normal distribution).
We find that the Bayesian evidence (logz) is 540.62 for the Free Gaussian model and is 541.22 for the \cotwo\ Gaussian.
Thus, there is no preference for the Free Gaussian (such as due to an anomalous noise spike) over a \cotwo\ Gaussian.
We encourage future observations with JWST to more robustly detect and constrain both \cotwo\ features and rule out noise at higher statistical significance.
}

\bibliographystyle{apj}
\bibliography{this_biblio}



\end{document}